\documentclass[10pt]{article}

\usepackage[a4paper, margin=1in]{geometry}
\usepackage{graphicx}
\usepackage{amsmath, amssymb}
\usepackage[round]{natbib}
\usepackage[colorlinks=true, linkcolor=blue, citecolor=blue, urlcolor=blue]{hyperref}
\usepackage{xcolor}
\usepackage{authblk}
\usepackage{setspace}
\usepackage{textcomp}
\usepackage{titlesec}
\usepackage{gensymb} 
\usepackage{subcaption}
\usepackage{float}
\usepackage{pdfpages}
\usepackage{caption}

\titleformat{\section}
  {\normalfont\bfseries\large}{\thesection}{1em}{}

\titleformat{\subsection}
  {\normalfont\bfseries\normalsize}{\thesubsection}{1em}{}

\titleformat{\subsubsection}
  {\normalfont\itshape\small}{\thesubsubsection}{1em}{}
\makeatletter
\renewcommand{\@biblabel}[1]{[#1]}
\makeatother

\title{\bfseries\Large Downscaling with AI reveals the large role of internal variability in fine-scale projections of climate extremes}

\author[1,2,3]{\normalsize Neelesh Rampal}
\author[1]{\normalsize Peter B. Gibson}
\author[2,3]{\normalsize Steven C. Sherwood}
\author[1]{\normalsize Laura E. Queen}
\author[1,4]{\normalsize Hamish Lewis}
\author[3]{\normalsize Gab Abramowitz}

\affil[1]{\small National Institute of Water and Atmospheric Research (NIWA), New Zealand}
\affil[2]{\small ARC Centre of Excellence for Weather of the 21st Century, University of New South Wales, Sydney, Australia}
\affil[3]{\small Climate Change Research Centre, University of New South Wales, Sydney, Australia}
\affil[4]{\small Te Aka Mātuatua School of Science, University of Waikato, Hamilton, New Zealand}

\date{}

\begin{document}
\maketitle
\vspace{-1em}
\begin{center}
\textcolor{blue}{\textbf{Corresponding author:} Neelesh Rampal (neelesh.rampal@niwa.co.nz)}
\end{center}
\begin{center}
\rule{\textwidth}{0.4pt}
\end{center}

\vspace{-1.5em}

\section*{Abstract}

The computational cost of dynamical downscaling limits ensemble sizes in regional downscaling efforts. We present a newly developed generative-AI approach to greatly expand the scope of such downscaling, enabling fine-scale future changes to be characterised including rare extremes that cannot be addressed by traditional approaches. We test this approach for New Zealand, where strong regional effects are anticipated. At fine scales, the forced (predictable) component of precipitation and temperature extremes for future periods (2080--2099) is spatially smoother than changes in individual simulations, and locally smaller. Future changes in rarer (10-year and 20-year) precipitation extremes are more severe and have larger internal variability spread than annual extremes. Internal variability spread is larger at fine scales that at the coarser scales simulated in climate models. Unpredictability from internal variability dominates model uncertainty and, for precipitation, its variance increases with warming, exceeding the variance across emission scenarios by fourfold for annual and tenfold for decadal extremes. These results indicate that fine-scale changes in future precipitation are less predictable than widely assumed and require much larger ensembles to assess reliably than changes at coarser scales. 

\section*{Introduction}
\doublespacing
An important limitation of Global Climate Models (GCMs) is their coarse resolution, which limits their ability to simulate climate changes at fine spatial scales \citep{maraun_precipitation_2010}. To address this limitation, dynamical downscaling is typically performed using Regional Climate Models (RCMs) \citep{giorgi_regional_1994}. RCMs are typically forced by boundary conditions from GCMs and run at higher spatial resolution over domains of interest \citep{giorgi_regional_1994}. RCMs generally add value over GCMs by more accurately simulating extremes and resolving regional climate processes \citep{aalbers_local-scale_2018, gibson_dynamical_2024, rummukainen_added_2016}(e.g. convection, orographic precipitation). Their high computational cost typically leads to small ensemble sizes, such that only a few studies have downscaled larger ensembles, which are typically restricted to individual GCMs and scenarios \citep{aalbers_local-scale_2018, kendon_variability_2023, leduc_climex_2019, poschlod_internal_2021, von_trentini_assessing_2019}. However, large ensembles of downscaled climate projections spanning many GCMs, scenarios and initial conditions, are needed to better sample a) overall climate variability and change, and b) extreme events \citep{maher_large_2021}. The latter is important as extreme events are, by definition, rare, and small ensembles may contain too few events to robustly estimate the forced climate response (signal) relative to internal variability (noise) \citep{aalbers_local-scale_2018}. 

Internal variability arises in multiple spatial and temporal scales, from chaotic mesoscale weather to decadal and planetary-scale climate patterns \citep{dai_decadal_2015, meehl_mechanisms_2012, zhang_can_2007}, and is an important source of uncertainty in global climate projections \citep{deser_uncertainty_2012, deser_projecting_2014, deser_insights_2020, lafferty_downscaling_2023, lehner_partitioning_2020,lehner_origin_2023, maher_quantifying_2020, martel_role_2018, rondeau-genesse_impact_2019}. It is distinct from model spread (variations from model to model) and scenario spread (from the range of possible anthropogenic influences such as CO2 emissions) \citep{hawkins_potential_2009, hawkins_potential_2011}. While internal variability averages out over the long term, it adds uncertainty to projections over any finite time period \citep{deser_uncertainty_2012}. Unlike the other uncertainty sources it is inherently unpredictable beyond a few years, so its uncertainty will not decrease with further research, model improvement, or clarity in emissions trajectory \citep{lehner_origin_2023}. Single Model Initial-Condition Large Ensembles (SMILEs) are a relatively new approach that more precisely isolates internal variability uncertainty from a GCM’s forced response, generated by running one GCM multiple times with perturbed initial conditions under constant forcing \citep{aalbers_local-scale_2018, bengtsson_can_2019, deser_uncertainty_2012, hawkins_irreducible_2016, machete_demonstrating_2016, von_trentini_assessing_2019}, with spread reflecting internal variability, and the ensemble average the forced response.   

The need to assess internal variability and extremes at finer scales calls for much larger ensemble sizes than have been possible so far for downscaling \citep{rampal_enhancing_2024}. Artificial intelligence (AI)-based RCM emulators are orders of magnitude faster than RCMs, offering an efficient way to generate large, downscaled ensembles \citep{chadwick_artificial_2011, doury_regional_2022, holden_emulation_2015, lopez-gomez_dynamical-generative_2025, rampal_enhancing_2024}. Recent studies have shown that AI-based emulators can capture historical climate (means, variability, extremes) and climate change signals for mean precipitation \citep{doury_suitability_2024, rampal_reliable_2025} well, but often underestimate future changes in extreme events, especially for precipitation \citep{addison_machine_2024, kendon_potential_2025, rampal_enhancing_2024}. Currently, no study has applied RCM emulators to large climate projection ensembles to investigate model, scenario, and internal variability uncertainty at fine scales. 

This study employs a generative-AI RCM emulator previously shown to accurately emulate the historical climate variability and warming-driven changes in mean and extreme precipitation \citep{rampal_reliable_2025, rampal_extrapolation_2024}. We demonstrate that this emulator, trained on a single RCM/GCM, accurately reproduces warming-driven RCM changes in mean and extreme daily maximum temperature (tasmax) and precipitation (pr) on driving fields from two other (previously unseen) GCMs, highlighting its out-of-sample generalizability. We use this emulator to downscale a large projection ensemble ($>$15,000 simulation years from 20 GCMs, 4 SSPs, including 2 SMILEs) for comprehensive regional uncertainty quantification. Unlike previous work focusing on seasonal or annual temperature or precipitation \citep{hawkins_potential_2009, hawkins_potential_2011, lehner_partitioning_2020, lehner_origin_2023}, this study investigates rare 10-year extremes. We show that internal variability remains a dominant source of uncertainty for these extremes.
\section*{Results}
\subsection*{Emulator Performance in Historical and Future Climates}

\begin{figure*}
    \centering
    \includegraphics[width=0.87\linewidth]{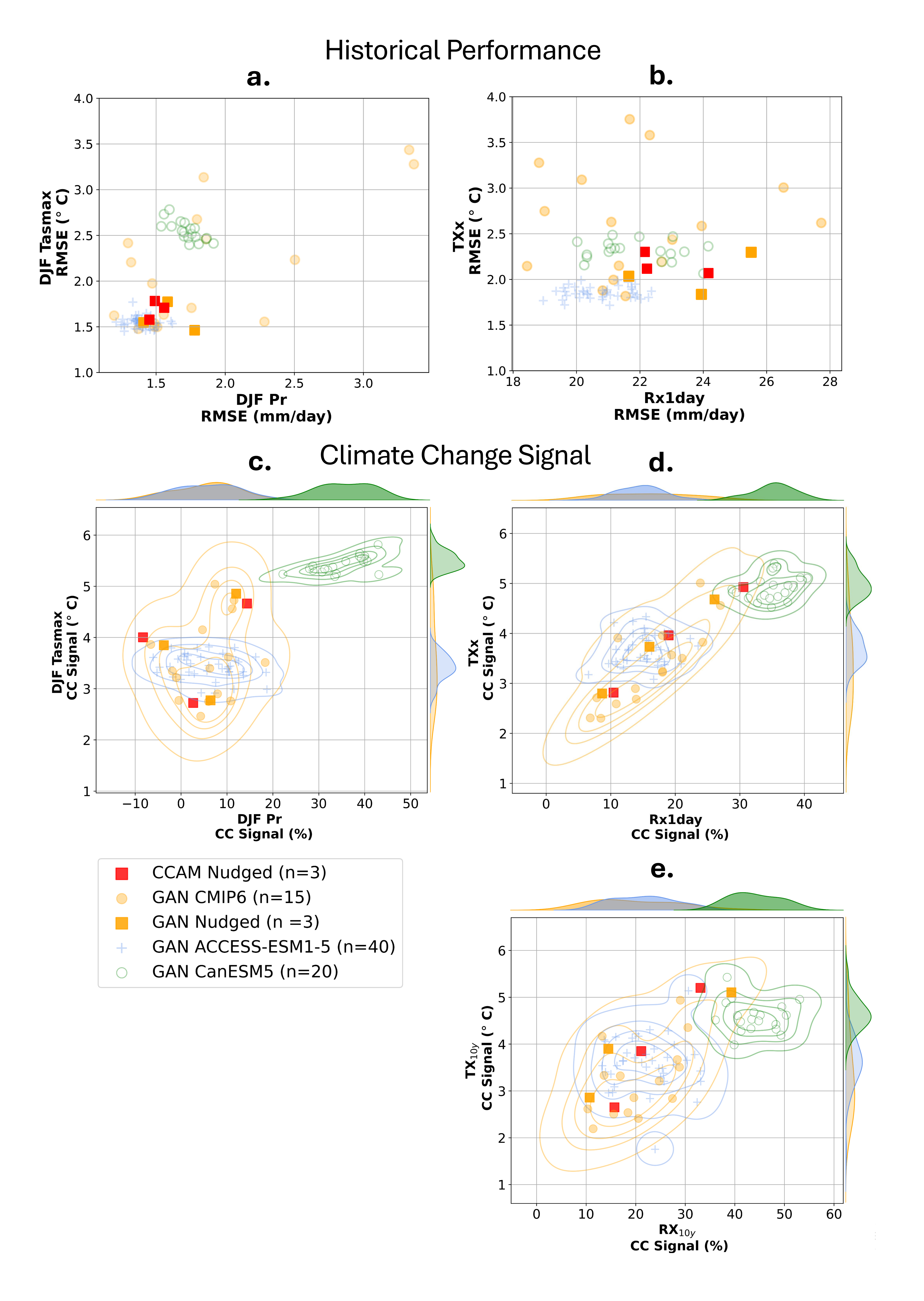}
    \caption{Performance, for historical climatology and future land-averaged climate change signals (SSP3-7.0), of an RCM emulator vs. RCM ground truth. (a,b): grid-cell level RMSE of downscaled historical (1986-–2005) daily maximum temperature (TXx, DJF Tasmax) and precipitation (Rx1day, DJF Pr) climatologies against the VCSN gridded observational dataset. (c-e): Climate-change signals, i.e., difference between the future (2080–2099) and the historical period (percentage changes for precipitation and absolute changes for temperature). GAN-downscaled simulations are represented for two Single Model Initial-condition Large Ensembles (SMILEs) – ACCESS-ESM1-5 (blue pluses, n=40) and CanESM5 (green open circles, n=20) – and for 15 other CMIP6 GCMs (orange filled circles, n=15). Simulations from the RCM driven by ACCESS-CM2, NorESM2-MM, and EC-Earth3 provide ground truth (red squares, n=3) directly comparable to corresponding GAN simulations driven by the same GCM data (orange squares, n=3).}
    \label{fig:Fig1C5}
\end{figure*}

To assess the emulator’s downscaling skill, we compare its historical performance, and its ability to represent climate change signals of means and extremes, to RCMs. We first compare historical climatologies (1986–-2005) of temperature and precipitation to the VCSN observational dataset (\autoref{fig:Fig1C5}a, b). The emulator performs similarly to the RCM (red vs. orange squares), showing similar root-mean-squared (RMS) error relative to observations (\autoref{fig:Fig1C5}a, b) for climatologies of summer December-February (DJF) mean and annual extreme temperature and precipitation (see Supplementary Fig. S1-S4 for spatial patterns of biases). Additionally, both the RCM and emulator show substantially reduced biases compared to simple interpolation from the fields of its driving GCM (Supplementary Fig. S5-S8), confirming the importance of downscaling. 

When the emulator is applied to the broader suite of CMIP6 GCMs (orange circles in \autoref{fig:Fig1C5}) and SMILEs (blue pluses, green circles)—for which direct RCM comparisons are unavailable—the skill shows considerable spread, with some performing as well as or better than the available RCMs while others perform significantly worse. Further analysis shows that while the emulator inherits some input biases from certain GCMs (e.g., a cold bias at $T_{500}$; Supplementary Fig. S9-S11), it generally improves over the GCM at finer scales (Supplementary Fig. S12–S14). Note that the GCMs selected for dynamical downscaling in prior work (red squares) were chosen based on criteria such as large-scale circulation biases, excluding many poorly performing models \citep{gibson_dynamical_2024, gibson_downscaled_2025}. In contrast, emulator-downscaled simulations assessed here include all GCMs, including those with larger biases stemming from the GCM, and are therefore less skilful on average, as expected.

We then evaluate the emulator’s ability to capture RCM climate change signals between the historical and the end-of-century (2080–-2099) periods in a high-emissions scenario (SSP3-7.0). These signals are assessed for the previous metrics but also for decadal extremes ($TX_{10y}$, $RX_{10y}$). When predictions are averaged over New Zealand land areas (\autoref{fig:Fig1C5}c-e), the emulator again effectively captures the spread amongst the three RCM simulations for mean and extreme changes. Spatial patterns and magnitudes also agree reasonably well for both variables (Supplementary Fig. S15-21). The regional warming rates (\autoref{fig:Fig1C5}c) for downscaled CMIP models are well correlated with the Equilibrium Climate Sensitivity (ECS) of the driving GCMs, indicating that the emulator preserves GCM-specific warming responses (Supplementary Fig. S22).

\subsection*{A range of possible futures for temperature and precipitation extremes}

Nationally, the emulator-downscaled expanded ensemble of CMIP6 models (\autoref{fig:Fig1C5}c–e, yellow symbols) shows a wider range of end-of-century vs historical land-averaged temperature and precipitation change signals than the three-member RCM ensemble. The spread of results in the two SMILEs (CanESM5; n=20, ACCESS-ESM1-5; n=40), measuring the impact of internal variability (\autoref{fig:Fig1C5}c-e), is generally larger for precipitation than temperature, consistent with earlier studies \citep{aalbers_local-scale_2018, leduc_climex_2019, von_trentini_assessing_2019}. For temperature, the range of changes across both SMILEs (difference between the hottest and coldest signals) is greater for annual and decadal extremes (\autoref{fig:Fig1C5}d–e) than for seasonal means (\autoref{fig:Fig1C5}c). This range is around 1$\degree$C for DJF tasmax and up to 4$\degree$C for $TX_{10y}$ in ACCESS-ESM1-5. As for precipitation, the spread remains large across all metrics—especially for $RX_{10y}$ and DJF mean precipitation.

\begin{figure*}
    \centering
    \includegraphics[width=1.0\linewidth]{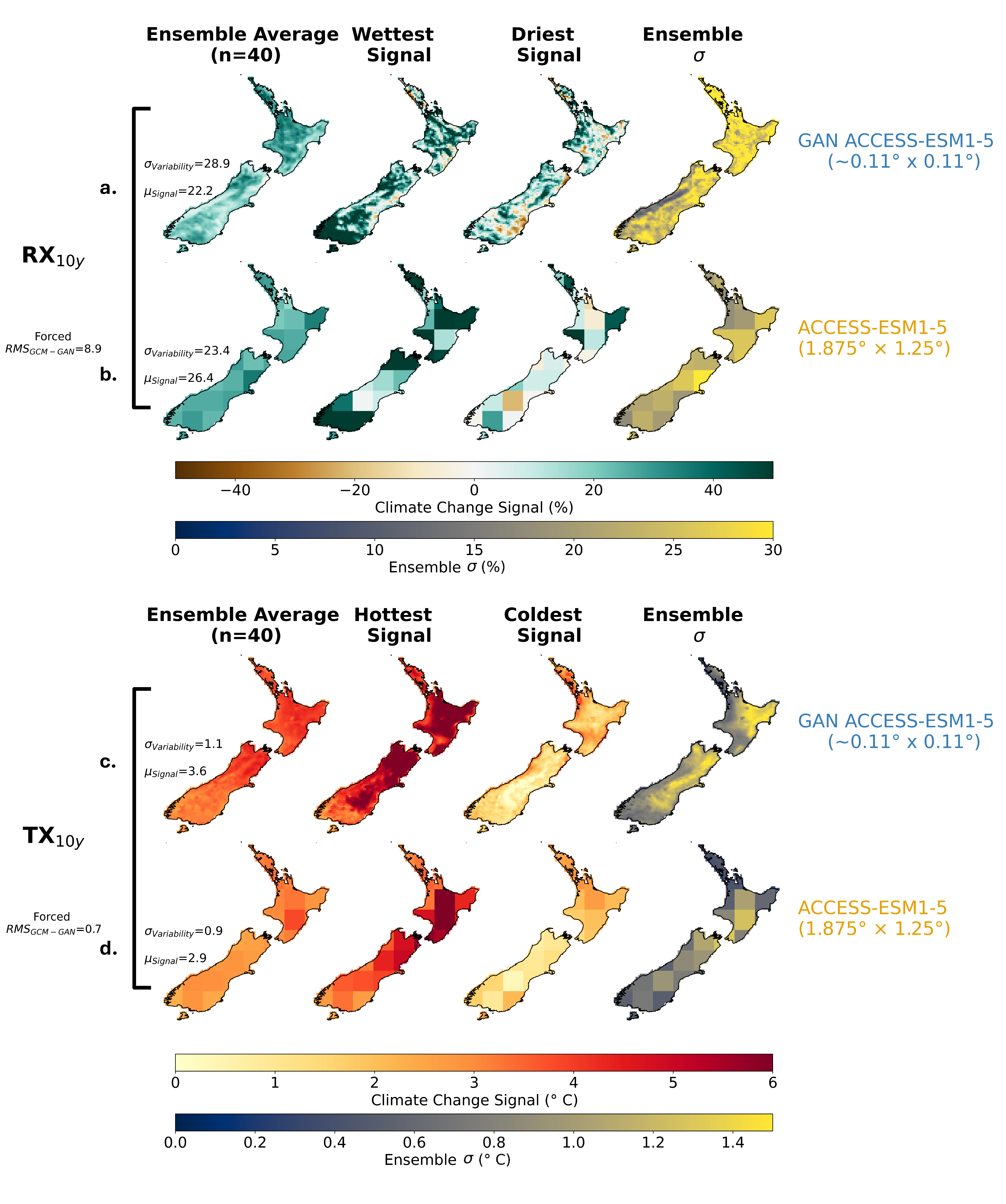}
    \caption{ Internal variability of 10-year extreme precipitation and temperature, at fine-scales (GAN) vs. from the original GCM. Climate change signals for extreme precipitation ($RX_{10y}$, a--b) and extreme temperature ($TX_{10y}$, c--d) from the GAN-downscaled ACCESS-ESM1-5 ensemble (top row in each block) vs. interpolation from the GCM (bottom row in each block). Columns show the ensemble mean, the wettest/hottest and driest/coldest signals, and the inter-member standard deviation ($\sigma$). The leftmost text quantifies key metrics: $\sigma_{\text{variability}}$ represents the spatially-averaged ensemble standard deviation, $\mu_{\text{signal}}$ is the spatially-averaged forced response (ensemble-mean), and the Forced RMS$_{\text{GCM-GAN}}$ denotes the Root Mean Square difference between the GAN-downscaled and original GCM’s forced responses.
}
    \label{fig:Fig2C5}
\end{figure*}

To isolate internal variability at fine scales, we focus on the ACCESS-ESM1-5 large ensemble spatial patterns of future changes in decadal extremes across individual ensemble members (wettest, driest, hottest, and coldest). The simulated changes in extremes from individual members of the emulator-downscaled ensemble (\autoref{fig:Fig2C5}) are generally noisy and unrepresentative of the smoother forced response obtained by averaging the whole ensemble (see Supplementary Fig. S23, for the CanESM5 large ensemble). This is particularly evident for $RX_{10y}$, where individual members show noisy spatial patterns of change that differ greatly in magnitude and even sign. To quantify the noise associated with internal variability (in a 20-year time frame), we compute the land-averaged signal-to-noise ratio (SNR = $\mu_{\text{signal}} / \sigma_{\text{variability}}$). For $RX_{10y}$, the SNR is 0.77 in the emulator versus 1.12 in the GCM, indicating that internal precipitation variability exceeds the climate-change signal at fine spatial scales, but not quite at coarse scales. In contrast, temperature shows higher and more consistent signal-to-noise ratios across the emulator and GCM (3.27 vs 3.22). Similarly, for annual extremes (RX1Day and TXx), individual members show large differences in spatial change patterns, though with less noise (Supplementary S24). Here, $\sigma_{\text{variability}}$ is smaller, leading to higher signal-to-noise ratios for both precipitation and temperature.

The climate change signals for $TX_{10y}$ and $RX_{10y}$ interpolated from individual ACCESS-ESM1-5 simulations (\autoref{fig:Fig2C5}b, d)—including the ensemble-mean forced response—lack the fine-scale geographical features shown by the emulator. Examples include less $RX_{10y}$ intensification on the east coast of both islands (which are generally drier), and generally more over the North Island than the South Island. For $TX_{10y}$, the downscaled fields generally warm more over the North Island that those of the GCM. Additionally, GCMs do not fully capture the spatial pattern of internal variability seen in the emulator-downscaled ensemble (\autoref{fig:Fig2C5}, last column), which shows distinct $RX_{10y}$ and $TX_{10y}$  spatial patterns that often shaped by orography (e.g., an east–west gradient in $TX_{10y}$). No such spatial pattern is observed in the GCMs, and the standard deviation is lower in most cases across both $RX_{10y}$ and $TX_{10y}$. The plausibility of these GAN-resolved patterns is supported by examining internal variability in the CanESM5 ensemble and historical interannual variability from RCM simulations and observations (Supplementary Figures S25-S27). 

The difference between the emulator and raw GCM forced responses can be interpreted as the emulator’s ‘potential added value’ over the GCM \citep{di_luca_potential_2013, di_virgilio_realised_2020}. Unlike conventional added value, which measures the reduction of historical biases relative to observations, potential added value compares how downscaling alters the climate change signal relative to its host GCM, and has been used in previous studies to assess individual RCMs \citep{di_luca_potential_2013, di_virgilio_realised_2020}. We quantify it as the RMS difference between the emulator and interpolated GCM forced responses (\autoref{fig:Fig2C5}; $RMS_{GCM-GAN}$). Although this RMS difference is modest relative to the overall signal (20–40\% for $TX_{10y}$  and $RX_{10y}$), it is comparable to internal variability in the case of $TX_{10y}$. Moreover, this suggests that downscaling can introduce signal differences similar in size to internal variability in some cases —an effect that could be treated as an additional type of model uncertainty that could be assessed in future work by for example comparing emulators trained on different RCMs \citep{rampal_enhancing_2024}.  

\subsection*{Assessing Confidence in Regional Precipitation Extremes Under High Emissions: A Case Study}

\begin{figure*}
    \centering
    \includegraphics[width=\linewidth]{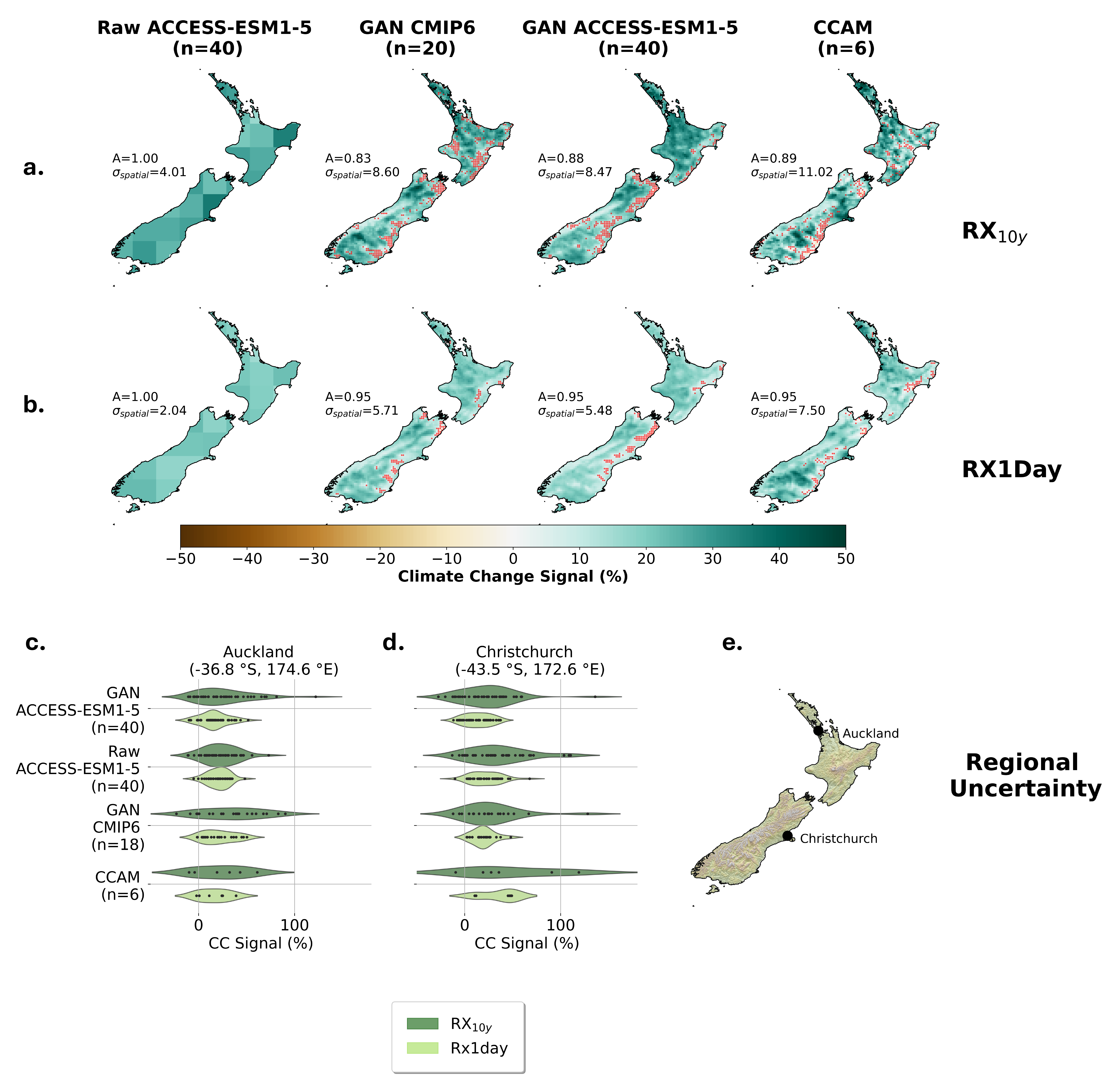}
    \caption{Spread in fine-scale extreme precipitation projections from a large ensemble: a comparison of GCM, RCM, and emulator. End of century (2080--2099) climate change signals (\%) for the SSP3-7.0 scenario relative to the historical climatology. (a--b): Ensemble-mean climate change signal in $RX_{10y}$ (a) and RX1Day (b) across four simulations—raw ACCESS-ESM1-5, GAN downscaled CMIP6 (20 members), GAN downscaled ACCESS-ESM1-5 (40 members), and CCAM (6 members). Red stippling is shown where fewer than 4/6 (66.6\%) members agree on the sign of the change. $\sigma_{\text{spatial}}$ shows the spatial standard deviation of the ensemble averages shown in (a) and (b), and A indicates the percentage of the area with sign agreement (regions without red stippling). Panels (c) and (d) display the spread of $RX_{10y}$ (dark green) and Rx1Day (light green) climate change signals across individual ensemble members for Auckland (c) and Christchurch (d), with locations shown in (e).}
    \label{fig:Fig3C5}
\end{figure*}

We present a case study (\autoref{fig:Fig3C5}) to demonstrate the value of emulator-based downscaling for capturing uncertainty in a societal decision-making context, focusing on how the ensemble spread (across models and internal variability) can be used to measure confidence in future projections (2080--2099 vs historical) of precipitation extremes (RX1Day, $RX_{10y}$). Overall, the emulator-downscaled ACCESS-ESM1-5 ensemble (\autoref{fig:Fig3C5}a, b; third column) shows more disagreement in the sign of change across ensemble members in eastern regions—particularly for $RX_{10y}$, which is absent in GCM (and is more spatially uniform). A similar pattern of disagreement is seen in a small CCAM ensemble (n=6) and across the GAN multi-model ensemble, further supporting the credibility of these signals and the GAN’s ability to represent uncertainty at fine scales. 

The bottom panel shows ensemble spread at the city scale (\autoref{fig:Fig3C5}c, d). For RX1Day, the emulator shows similar spread across multi-model and internal variability ensembles (GAN ACCESS-ESM1-5), with signals differing by up to 50\% between the wettest and driest members. $RX_{10y}$ shows even greater spread (–30\% to +150\% in Christchurch; \autoref{fig:Fig3C5}). Interestingly, over Christchurch, the CCAM ensemble (n = 6) forms two distinct clusters, with highlighting the sampling limitations of small ensembles compared to the larger emulator ensemble. 

\subsection*{Uncertainty decomposition at fine scales}
\begin{figure*}
    \centering
    \includegraphics[width=0.85\linewidth]{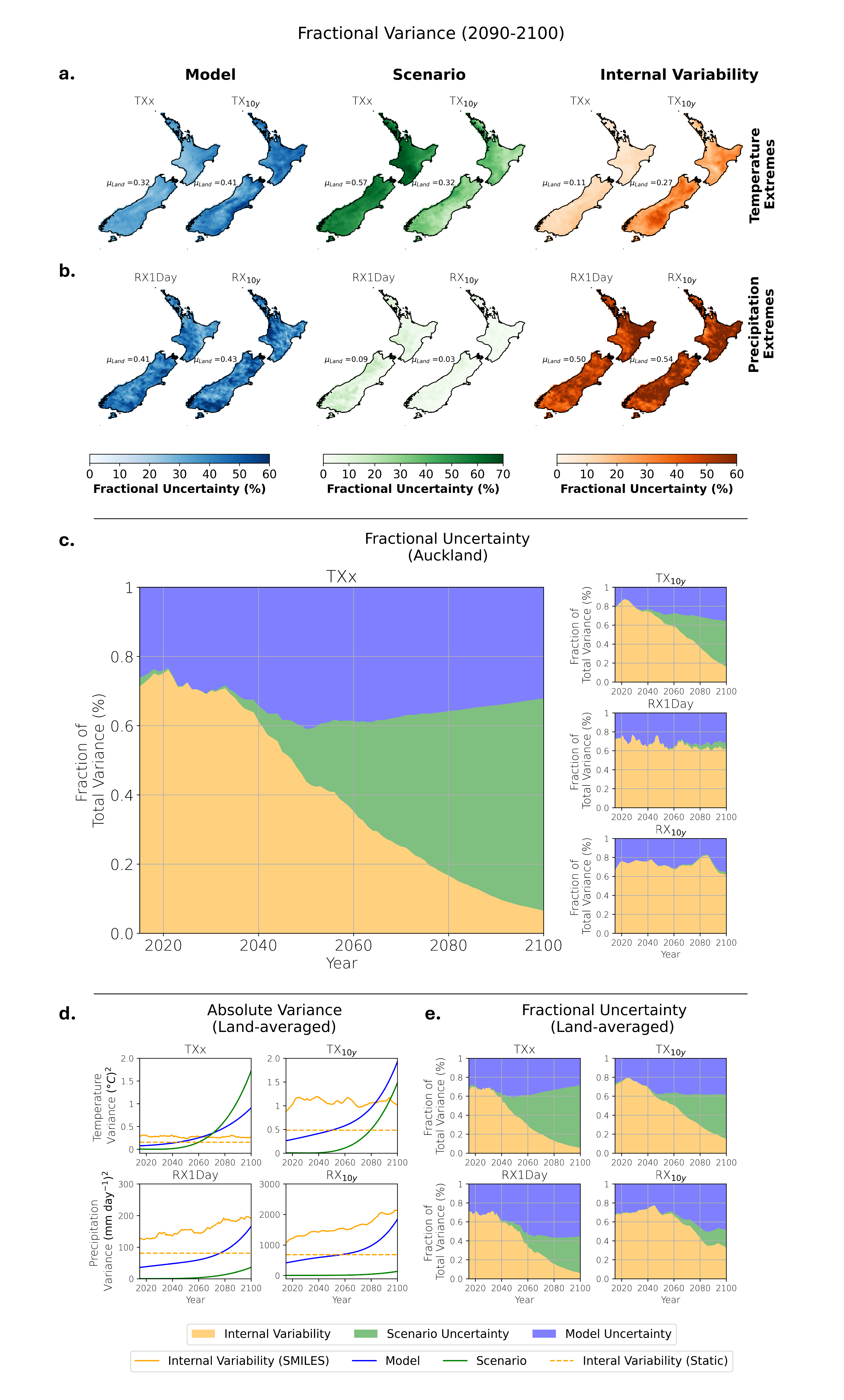}
    \caption{Temporal and spatial variability of uncertainty decomposition for extreme temperature and precipitation. (a-b) spatial patterns of fractional uncertainty for temperature (a) and precipitation extremes (b), from model spread (left column), scenario spread (middle), and internal variability (right). Within each column, annual extremes (TXx, RX1Day) are on the left and decadal extremes ($TX_{10y}$,$RX_{10y}$) on the right. (c) Temporal evolution of fractional uncertainty for a grid cell over Auckland (-36.8$\degree$S, 174.6$\degree$E) for all extreme metrics. (d) Absolute variance, averaged across all grid points, for each uncertainty source across TXx, $TX_{10y}$, RX1Day, and $RX_{10y}$. (e) Same as (c) but computes fractional contributions after first land-averaging (or coarsening) the annual and decadal extremes. }
    \label{fig:Fig4C5}
\end{figure*}

Following previous studies \citep{hawkins_potential_2009, lehner_partitioning_2020}, we decomposed the time-evolution of total variance (interpreted here as uncertainty) in projections of annual and decadal extremes into contributions from the climate model (hereafter model), scenario and internal variability (\autoref{fig:Fig4C5}a, b). Annual and decadal extremes are smoothed using a 10-year rolling average to reduce noise \citep{hawkins_potential_2009, lehner_partitioning_2020}, so the value for 2099 represents the average from 2090 to 2099. In 2015–2024, internal variability accounts for over 70\% of variance in both annual and decadal temperature and precipitation extremes (Supplementary Fig. S29). In 2090–2099, internal variability plays a smaller role in the uncertainty of annual ($\sim$11\% on-average) and decadal ($\sim$27\% on-average) temperature extremes, with scenario spread becoming dominant, followed by model spread (\autoref{fig:Fig4C5}a). As for annual and decadal precipitation extremes, the fractional contribution of internal variability decreases slightly, though it remains the largest source of uncertainty (averaging over 50\%).

The uncertainty breakdown varies regionally due to orography. By the end of the century, temperature extremes exhibit $\sim$20\% more internal variance on the east coast than the west on both islands – which are separated by mountain ranges. A similar but less pronounced pattern is seen for precipitation. Previous studies using GCMs have identified regional differences in internal variability \citep{blanusa_internal_2023, gibson_storylines_2024, hawkins_potential_2009, lafferty_downscaling_2023, lehner_partitioning_2020, yip_simple_2011} albeit not at the fine scales explored here. Those studies indicated that mid-latitude regions, such as New Zealand, typically exhibit greater internal variability in decadal-mean precipitation than do tropical regions, though this difference is smaller for temperature \citep{blanusa_internal_2023, lehner_partitioning_2020}.  

We then compare fractional importance of internal variability at individual grid-cells (i.e. Auckland) to its importance for the land average (where projected changes are averaged before computing contributions). For temperature extremes, internal variability is only slightly less important for the land average than for specific locations (\autoref{fig:Fig4C5}c vs \autoref{fig:Fig4C5}e). In contrast, land-averaged precipitation extremes (e.g., RX1Day, $RX_{10y}$) show significantly less internal variability. This leaves a greater fractional contribution from scenario spread for precipitation extremes: spatial averaging, even over a small country like New Zealand, will "average out" fine-scale processes and reduce the role of internal variability. This shows that internal variability is greater at finer scales and will likely depend on the RCM resolution. This aligns with prior research comparing global averages and individual grid points in GCMs for decadal precipitation means \citep{deser_insights_2020}.

By the end-of-century (2090-2100), internal variability contributes more to total uncertainty in rarer decadal extremes than in annual temperature extremes, as evident in the spatial patterns (\autoref{fig:Fig4C5}a, b) and temporal evolution (\autoref{fig:Fig4C5} bottom two rows). While internal variability is generally the smallest source of uncertainty for temperature extremes by 2090-2100, its contribution remains non-negligible. In comparison studies using GCMs, have often found this uncertainty to be negligible by 2090-2100 for decadal temperature means \citep{hawkins_potential_2009, lehner_partitioning_2020}. For precipitation, we show that internal variability is the main source of uncertainty for seasonal (Supplementary Fig. S30), annual, and decadal extremes, showing similar importance across all metrics, consistent with previous studies using GCMs \citep{blanusa_internal_2023, lafferty_downscaling_2023, lehner_partitioning_2020}.

\begin{figure*}
    \centering
    \includegraphics[width=0.75\linewidth]{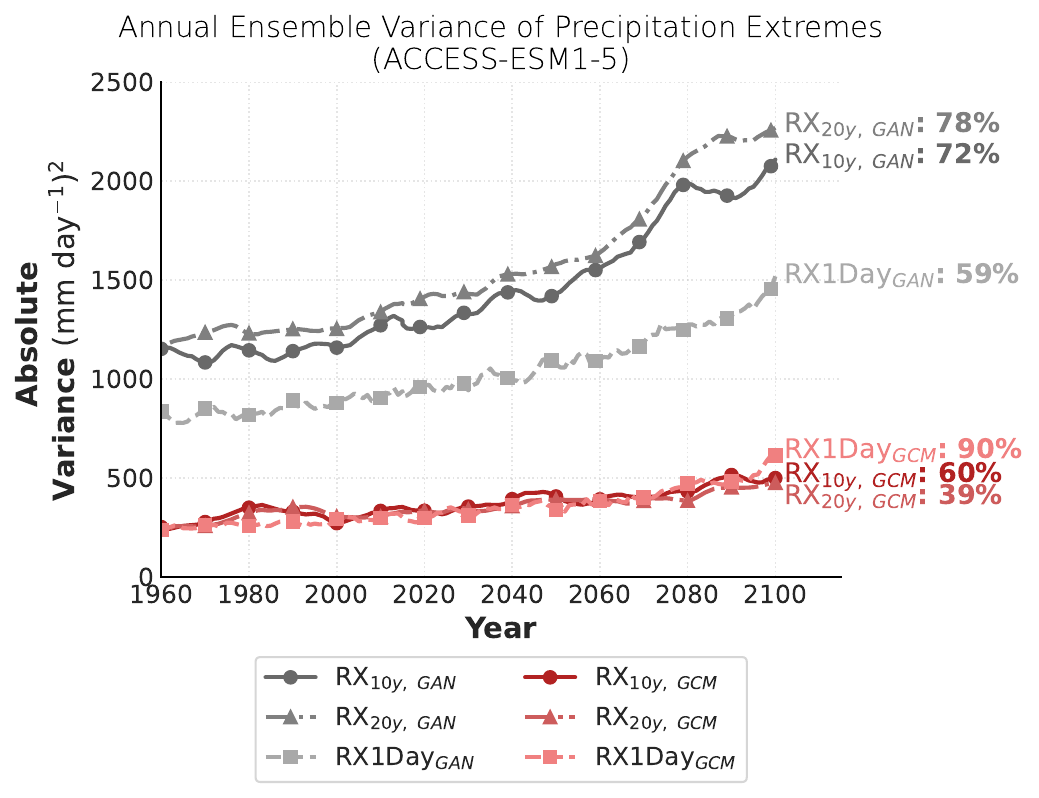}
    \caption{Temporal evolution of internal variability variance for precipitation extremes (RX10y, RX20y, RX1Day) in the raw vs GAN-Downscaled ACCESS-ESM1-5 ensemble. Here, internal variability variance is computed annually (no decadal smoothing as in \autoref{fig:Fig4C5}d), comparing GAN (grey) and raw GCM (red) precipitation. Variance increases are shown in text for the future period (2080–2100) relative to the historical baseline (1986–-2005).}
    \label{fig:Fig5C5}
\end{figure*}

Many studies assume that internal variability is constant in time, and typically estimate its variance using multi-model GCM ensembles \citep{blanusa_internal_2023, hawkins_potential_2011, lafferty_downscaling_2023}, rather than SMILEs as done here. We show that the land-average internal variance of fine-scale annual and decadal precipitation extremes increases over time, shown by positive trends in absolute ensemble variance in \autoref{fig:Fig4C5} (bottom row, lefthand; yellow line). This variance is larger and increases more for rarer extremes than for annual extremes (\autoref{fig:Fig4C5}d, \autoref{fig:Fig5C5}). The GAN-downscaled ACCESS-ESM1-5 ensemble shows increases of 59\% for annual, and 72\% and 78\% for 10- and 20-year extremes, respectively, in 2080-–2099 relative to 1986–-2005 (\autoref{fig:Fig5C5}). For the raw ACCESS-ESM1-5 ensemble, the variance remains similar across annual, 10-year, and 20-year extremes. In fact, rarer extremes show comparatively smaller increases in variance. Overall, this increase in variance produces no trend in fractional uncertainty (\autoref{fig:Fig4C5}c, e)—due to concurrent increases in model and scenario variance—it contrasts with studies assuming static internal variability, which would imply a reduction in its relative contribution over time. We find no such reduction. By 2100, the SMILE-based estimates of fine-scale internal variability are over two times larger than those derived from a static assumption.

The positive trends in ensemble variance are likely linked to increases in evaporation and atmospheric moisture in a warmer climate \citep{fischer_observed_2016}, which has been used to explain increased precipitation variance in both observations and future projections across annual to decadal timescales \citep{de_vries_increasing_2024, pendergrass_precipitation_2017, schwarzwald_changes_2021, wood_changes_2021, zhang_anthropogenic_2024, zhang_constraining_2022}. For temperature extremes, we do not see any distinct trends in the magnitude of internal variability, though SMILE-based estimates are generally larger than those from static methods, consistent with a previous study \citep{lehner_partitioning_2020}. 

\section*{Discussion and Conclusion}
A common concern regarding AI downscaling algorithms, particularly in the context of extreme events and climate change, is their tendency to fail to generalize beyond the training data \citep{addison_machine_2024, doury_suitability_2024, kendon_potential_2025}. But here, we show that the generative-AI RCM emulator is more robust to these criticisms, accurately capturing climate change signals for both mean and rare extremes from GCMs not seen during training. This massive increase in ensemble size over previous methods finally enables a clear view of the fine-scale forced response to climate change and the relative contributions to future uncertainty from models, scenarios, and internal variability at fine scales, as well as a more quantitative analysis of rare climate extremes (10-year).

The spread from (unpredictable) internal variability in 2080–2099 projections can exceed the forced response for precipitation extremes at fine spatial scales and is more pronounced at fine scales than at those accessible to the driving GCMs. It contributes over 50\% of the total uncertainty (variance) in annual and decadal (10-year) extremes over New Zealand—exceeding model and scenario contributions. Internal variability plays a smaller role in temperature extremes at fine scales than for precipitation, contributing $\sim$11\% of annual and $\sim$27\% of decadal uncertainty in projections, but is not negligible as sometimes assumed \citep{hawkins_potential_2009, lehner_partitioning_2020}.

Because of the large internal variability at fine scales, downscaled projections from individual members (i.e. wettest and driest members) of a climate model provide only limited insight into the model’s true forced response \citep{aalbers_local-scale_2018, deser_uncertainty_2012, deser_insights_2020, maher_large_2021}, which is spatially smoother. This is especially true for rare extremes where fine-scale internal variability can obscure the forced signal \citep{kendon_variability_2023}. While the model’s forced response under a given scenario indicates the expected outcome, individual ensemble members illustrate the range of plausible future outcomes that may occur as result of unpredictable internal variability, as only one future will ultimately unfold.

Internal variability is more pronounced at fine spatial scales than for averages over GCM grid cells or a region the size of New Zealand, especially for precipitation extremes. This means fine-scale precipitation changes are less predictable than regional averages. Moreover, while we can more accurately predict a regional average, or likelihood of an extreme event somewhere within the region, we cannot anticipate exactly where it will occur. These findings confirm that coarse-resolution models underestimate the influence of internal variability at local scales, as a suggested by a previous RCM-SMILE study \citep{aalbers_local-scale_2018}.

Lastly but importantly, we find that internal variability variance in precipitation extremes roughly triples by the end of the century, in line with spread contributions from other sources, but well exceeding the increase in precipitable water due to warming. It is also nearly three times larger than static estimates of variance from multi-model ensembles. A previous study also reported such increases \citep{blanusa_internal_2023}, though smaller than those observed here. Studies using multi-model ensembles to assess future uncertainty need to take this into account, as well as the uncertainty in internal variability itself as shown by variations among different SMILEs.

Overall, we show that one cannot assume future changes in fine-scale precipitation extremes are predictable; they are inherently unpredictable, and only limited amount of information can be inferred about their changes. This finding for New Zealand is likely applicable to other regions. New Zealand’s orographically influenced climate may allow more predictable fine-scale changes than flatter regions, making it a useful baseline for assessing predictability elsewhere.

To better account for this uncertainty, societal decision-making and climate adaptation efforts may require a more nuanced approach \citep{harrington_procurement_nodate}, calling for on the one hand, significantly larger downscaled ensembles than previously possible; and on the other, risk assessments that accept the limitations of what can be predicted at local scales. RCM emulators offer a promising and inexpensive way to leverage RCMs to generate these large ensembles, allowing for better sampling of rare extremes and internal variability. When doing so, however, emulators require careful design and evaluation to ensure that they are fit-for-purpose and can reliably extrapolate beyond their training data.

\section*{Methods}
\subsection*{Predictor and Target Variables}
We use a deep learning-based emulator adapted from a previous study \citep{rampal_extrapolation_2024, rampal_reliable_2025} to separately downscale daily precipitation (pr; [mm/day]) and daily maximum temperature (tasmax; K) over the New Zealand region (165$\degree$E–184$\degree$W, 33$\degree$S–51$\degree$S). The emulator was trained on predictor and target variables from the Conformal Cubic Atmospheric Model (CCAM), a global non-hydrostatic model with a variable-resolution grid widely used for downscaling \citep{chapman_evaluation_2023, gibson_dynamical_2024, thatcher_using_2009}. Further information regarding CCAM's performance for this region is evaluated in detail in previous studies \citep{campbell_comparison_2024, gibson_downscaled_2025, gibson_dynamical_2024}.

The daily precipitation target variable is logarithmically normalized, whereas temperature is normalized relative to its spatio-temporal mean and variance (single value for $\mu_t$ and $\sigma_t$) to preserve spatial gradients. We use the same daily-averaged coarsened CCAM predictors at 1.5$\degree$—u,v,t, and q at 500 and 850 hPa—as in previous studies \citep{rampal_reliable_2025, rampal_extrapolation_2024}. This is commonly referred to as the perfect framework setup \citep{doury_regional_2022, rampal_enhancing_2024}. For precipitation, predictors are normalized using the spatio-temporal mean and standard deviation from the full training data, consistent with previous studies on this model \citep{bailie_quantile-regression-ensemble_2024, rampal_high-resolution_2022, rampal_reliable_2025, rampal_extrapolation_2024}. For temperature, we use a slightly different approach. We normalized daily predictor variables by standardizing each based on its daily mean and standard deviation; this approach was found to perform better than the normalization used for precipitation temperature in a previous study \citep{doury_regional_2022, doury_suitability_2024}. These daily means and standard deviations are also used as predictors (normalized by their temporal means).

The perfect-model setup refers to using the coarsened CCAM large-scale circulation fields as predictors (instead of using the GCM large-scale fields directly; imperfect framework). Training an emulator through the imperfect framework is more challenging as the RCM's mean state can significantly deviate from the GCM \citep{bano-medina_transferability_2024, bartok_projected_2017, boe_simple_2023, doury_suitability_2024, sorland_bias_2018}, and the relationships learned are often less portable than the perfect framework \citep{bano-medina_transferability_2024, boe_simple_2023, hernanz_critical_2022, rampal_enhancing_2024}. Our results indicate that emulators trained in the perfect framework perform well when applied directly to GCM inputs, with only slightly larger errors than when applied in the perfect setup (i.e., to coarsened CCAM fields; See Supplementary Fig. S32-S35 for more information). Both emulators are trained on 140 years (1960–2100) of RCM simulations forced by ACCESS-CM2 GCM. We also experimented with training on multiple GCMs (three or five), which did not improve in or out-of-sample performance (see Supplementary S36-S37).

\subsection*{RCM Emulator Architecture}

The RCM emulator architecture is a residual GAN adapted from previous studies \citep{rampal_reliable_2025, rampal_extrapolation_2024} and consists of two components. First, a deterministic Convolutional Neural Network is trained to emulate a specific variable (i.e. precipitation), which captures the predictable, large-scale component of precipitation driven by regional circulation. Then, a GAN is trained on the residuals—the differences between the CNN output and the RCM truth, which allows the GAN to better extrapolate to warmer climates and more accurately downscale extreme events \citep{rampal_reliable_2025, rampal_extrapolation_2024}. Thus, our approach is a “conditional GAN (cGAN)”, though referred to for simplicity as GAN.

The deterministic CNN algorithm is based on the U-Net architecture. The U-Net architecture features both contracting and expansive pathways, where some intermediate layers are "skip-connected" between the pathways. The U-Net architecture incorporates residual blocks within both contracting and expansive pathways. In this study, we have incorporated several additional residual blocks and increased the number of filters used in each block compared to \citep{rampal_reliable_2025, rampal_extrapolation_2024}, resulting in 3.5 million trainable parameters. Inputs to the deterministic CNN include the large-scale predictors (u,v,q,t at 850 and 500hPa) and topography. 

GANs involve two competing models: a generator and a critic, which are trained to compete with each other. The generator aims to produce downscaled precipitation that the critic cannot distinguish from real CCAM simulations, while the critic learns to differentiate between real and generated data \citep{goodfellow_generative_2014, mirza_conditional_2014}. This adversarial training framework encourages the generator to learn a loss function that goes beyond grid-cell differences, promoting outputs that are structurally and distributionally consistent with the real data \citep{annau_algorithmic_2023, glawion_spategan_2023, miralles_downscaling_2022, rampal_reliable_2025}. The GAN architecture is identical to the U-Net’s and is trained to predict the residual between the deterministic CNN's output and the ground truth RCM. It includes one additional input: the U-Net's prediction is fed directly into the generator. The GAN is trained using a composite loss:
\[
G_{\text{loss}} = \text{MSE}(y_{\text{true}}, \hat{y}_{\text{pred}}) + \lambda_{\text{adv}} G_{\text{loss,adv}} + IC(y_{\text{true}}, y_{\text{pred}}),
\]

where \(\lambda_{\text{adv}} = 0.01\) controls the weight of the adversarial loss (\(G_{\text{loss,adv}}\)), and \(G_{\text{loss,adv}} = -\overline{D(y_{\text{pred}})}\) is the negative mean critic score. We adapt the intensity constraint (IC) by applying regionally pooled maxima over 20-pixel ($\sim$350 km) windows—rather than a global maximum as in a previous study \citep{rampal_reliable_2025}—to better capture regional variability in extreme precipitation. Previous studies provide a comprehensive evaluation of how these training parameters affect model performance \citep{rampal_reliable_2025, rampal_extrapolation_2024}.

Following previous studies \citep{gulrajani_improved_2017, leinonen_stochastic_2021}, both the generator and discriminator are trained using the Adam optimizer with an initial learning rate of \(2 \times 10^{-4}\) and a batch size of 32. The deterministic baseline uses a learning rate of \(7 \times 10^{-4}\). All models are trained for 200 epochs ($\sim$24 hours) on two NVIDIA A100 GPUs (80 GB RAM). Learning rate decay helps stabilize GAN performance, with similar results observed using 180 instead of 200 epochs.

\subsection*{Generating a Large Ensemble of Climate Projections}
We selected 20 GCMs covering the historical period (1960–2014) and four SSPs (SSP1-2.6, 2-4.5, 3-7.0, 5-8.5; 2015–2099) to capture both model structural and scenario uncertainty (see Fig. S39 for a list of GCMs selected). We also included two SMILEs, CanESM5 (n=20) \citep{swart_canadian_2019} and ACCESS-ESM1-5 (n=40) \citep{ziehn_australian_2020}—for the historical and SSP3-7.0 periods, as studies often average the spread from multiple SMILEs to account for GCM-dependent differences and compare it to model and scenario uncertainty \citep{lehner_partitioning_2020}. GCMs were selected based on the availability of daily large-scale predictors across all scenarios on the Earth System Grid Federation (ESGF), which constrained model choice due to data completeness (excluding SMILE simulations). 

For downscaling with the residual GAN RCM emulator, each GCM-scenario input was normalized using the mean and variance of the training data (CCAM-coarsened ACCESS-CM2) and passed through the deterministic CNN and residual GAN independently, enabling parallel inference across GCMs and scenarios. The Residual GAN output was added to the deterministic baseline and unnormalized (e.g., \( pr = \exp(z), \; T = z \times \sigma_t + \mu_t \)) to generate daily precipitation and maximum temperature. For each day, a single GAN prediction is generated from a randomized noise vector, rather than using an ensemble as in previous studies \citep{rampal_reliable_2025}, since variability was found to be minimal for the statistics explored here. Each day is downscaled independently. No bias correction was applied to maintain consistency with conventional RCM workflows. Downscaling one simulation takes approximately 3 minutes (historical) or 4 minutes (scenario) on an A100 GPU, totaling approximately 6 hours to generate the entire dataset using 4 A100 GPUs. We did not apply bias correction to the GCM inputs to be consistent and comparable with conventional numerical RCMs.

\subsection*{Historical Evaluation of RCM Emulators}
The performance of the Residual GAN is assessed through 1) comparison with ground truth RCMs and (2) evaluation against observational data.
\subsubsection*{Evaluation against dynamical simulations}
To evaluate how well the emulator reproduces the statistical and physical characteristics of the RCM, we compare the emulator trained on ACCESS-CM2 against two unseen (from training) ground truth RCM simulations (EC-Earth3 and NorESM2-MM). We focus on future climate change signals in the main text (\autoref{fig:Fig1C5}c–e) but also compare historical performance (relative to ground truth RCMs) in Supplementary Fig. S32–S35.

\subsubsection*{Evaluation against observations}
Evaluation against observations (\autoref{fig:Fig1C5}a-b) is performed using the Virtual Climate Station Network (VCSN) reference observational dataset, comprising $\sim$700 temperature stations and $\sim$1300 rain gauges across New Zealand at $\sim$5 km resolution \citep{tait_thin_2006, tait_assessment_2012}. Metrics are computed and then interpolated to the 5km VCSN grid using bilinear interpolation. This evaluation is essential for assessing emulator skill on out-of-sample GCMs, where no ground truth RCM simulations are available for comparison. If the emulator's historical errors on out-of-sample GCMs are similar to those on GCMs with available RCM ground truth, it suggests the emulator generalizes well \citep{rampal_enhancing_2024}.

\subsection*{Internal Variability in Future Climate Projections}
To qualitatively illustrate internal variability and model uncertainty in climate change responses, we compute temperature and precipitation signals for the SSP3-7.0 scenario across a broad range of GCMs (n=18 plus two SMILEs) using the same historical and future periods as defined above. We also examine signals of decadal maxima, computed as the average maximum across two decades spanning each climatological period. 

\subsubsection*{Fractional Uncertainty in Climate Projections}
To quantify the contributions of different sources of uncertainty in climate projections, we decompose the total variance (T) at each grid point and year into internal variability (I), model uncertainty (M), and scenario uncertainty (S), following previous studies \citep{hawkins_potential_2009, lehner_partitioning_2020}. While this method assumes M, S, and I are independent—an assumption not strictly valid—previous work shows that accounting for their covariance has minimal impact on the results \citep{yip_simple_2011}. We assume this holds true at finer scales. We compute the fractional contribution of each variance component by normalizing with total variance (e.g., I’ = I/T). This decomposition is applied to extreme precipitation ($RX_{10y}$  ,RX1Day) and temperature (TXx, $TX_{10y}$).

We first compute anomalies in annual extremes (RX1Day, TXx) relative to 1986–-2005 climatology (percent change for precipitation, absolute for temperature). Note: We use absolute precipitation anomalies as a pose to previous studies using percentage changes, but find similar results between the two approaches (see Supplementary Figure S28 for a comparison). For annual extremes, the forced response is estimated by fitting a second-degree polynomial (1950–2100) to rolling decadal means of anomalies. For decadal extremes ($RX_{10y}$, $TX_{10y}$), decadal maximums are computed using a 10-year rolling window, smoothed with a subsequent 10-year rolling average, and then fitted with a second-degree polynomial. Model uncertainty is defined as the variance across GCM-specific forced response fits (see Supplementary S39-s40 for an illustration of the GCM-specific fits), averaged over scenarios. Scenario spread is quantified as the variance across the unweighted, multi-model average (forced) response fits for each scenario. We estimate internal variability using two methods: (1) the variance of residuals from a polynomial fit, averaged across models(Hawkins \& Sutton, 2009) (shown as static variability only in \autoref{fig:Fig4C5}d), and (2) the variance across SMILE ensemble members \citep{lehner_partitioning_2020} (used in \autoref{fig:Fig4C5}a–e). For (2), the final estimate is the average from two SMILEs (For an illustration of precipitation variance, see Figure S31 for CanESM5 and Figure 5 for ACCESS-ESM1-5). The first method gives a static estimate of internal variability, while the second provides a time-varying estimate and is used by default unless stated otherwise. An important limitation of the first method—and of any approach that relies on a single realisation per global model—is that it conflates model uncertainty with internal variability and assumes internal variability is constant over time \citep{lehner_partitioning_2020}.

In \autoref{fig:Fig4C5}, we show land-averaged absolute variance (\autoref{fig:Fig4C5}d) and land-averaged fractional uncertainty (\autoref{fig:Fig4C5}e), which differ in their calculation. The former illustrates the non-stationarity of variance, while the latter shows how land-averaging smooths internal variability in trends and projections. For \autoref{fig:Fig4C5}d, we calculate model, scenario, and internal variances at each land grid point and then average the absolute variances across all land points (unweighted). Second (\autoref{fig:Fig4C5}e), we land-averaged future projections/trends (e.g., RX1Day anomalies at a given year), then calculate the variances and fractional contributions from model, scenario, and internal sources. In \autoref{fig:Fig5C5}, ensemble variance is calculated annually without decadal smoothing, unlike \autoref{fig:Fig4C5}d.

%


\subsection*{Acknowledgments}
N.R P.B.G and L.Q received funding from the New Zealand MBIE Endeavour Smart Ideas Fund (C01X2202). N.R and S.C.S acknowledge the support of the Australian Research Council 21st Century Weather (CE230100012). S.S, N.R and P.B.G would also like to acknowledge support from MBIE Strategic Science Investment Fund (SSIF). The authors would also like to acknowledge the New Zealand eScience Infrastructure for providing access to GPUs.

\subsection*{Data and Code Availability}
The ERA5 reanalysis data used in this study are available for download from the Climate Data Store (Hersbach et al., 2023). CMIP6 data used in this study, were at the time of submission, available from the Earth System Grid Federation (ESGF) archive: https://esgf-node.llnl.gov/projects/cmip6/.

The RCM emulator code and datasets supporting this study are available on \href{https://zenodo.org/records/13755688}{Zenodo}. The code for training the RCM emulator is available at: \href{https://github.com/nram812/On-the-Extrapolation-of-Generative-Adversarial-Networks-for-downscaling-precipitation-extremes}{https://github.com/nram812/On-the-Extrapolation-of-Generative-Adversarial-Networks-for-downscaling-precipitation-extremes}. The code for generating the figures and the large ensemble of climate projections used in this study—including CMIP6 data downloading and preprocessing scripts—is available at: \href{https://github.com/nram812/Downscaling-with-AI-reveals-large-role-of-internal-variability-in-fine-scale-projections}{https://github.com/nram812/Downscaling-with-AI-reveals-large-role-of-internal-variability-in-fine-scale-projections}.



\renewcommand\refname{References}
\bibliographystyle{apalike}
\textnormal{\bibliography{references}}
\newpage





\section*{Supplementary material}

Supporting Information for

# Downscaling with AI enables sampling of fine-scale internal variability in regional climate projections

Neelesh Rampal[1,3,4], Peter B. Gibson[1], Steven Sherwood[3,4], Laura Queen[1], Hamish Lewis[1,2], Gab Abramowitz[4]

[1] National Institute of Water and Atmospheric Research (NIWA), New Zealand
[2] Te Aka Mātuatua School of Science, University of Waikato, Hamilton, New Zealand
[3] ARC Centre of Excellence for Weather of the 21st Century, University of New South Wales, Sydney, Australia
[4] Climate Change Research Centre, University of New South Wales, Sydney, Australia

# Overview

This document contains the supporting information for the manuscript titled: *Downscaling with AI reveals large role of internal variability in fine-scale projections of climate extremes*. The contents (Figures S1-S40) are divided into two sections as outlined below with several subsections.

# Contents

1

# Results

## Emulator Performance in Historical and Future Climates

This section contains figures related to the above section in the main manuscript. We first begin by focusing on performance the historical period of simulation, where evaluation is performed against the VCSN observational dataset.

### *Historical Evaluation (Figures S1-S13)*

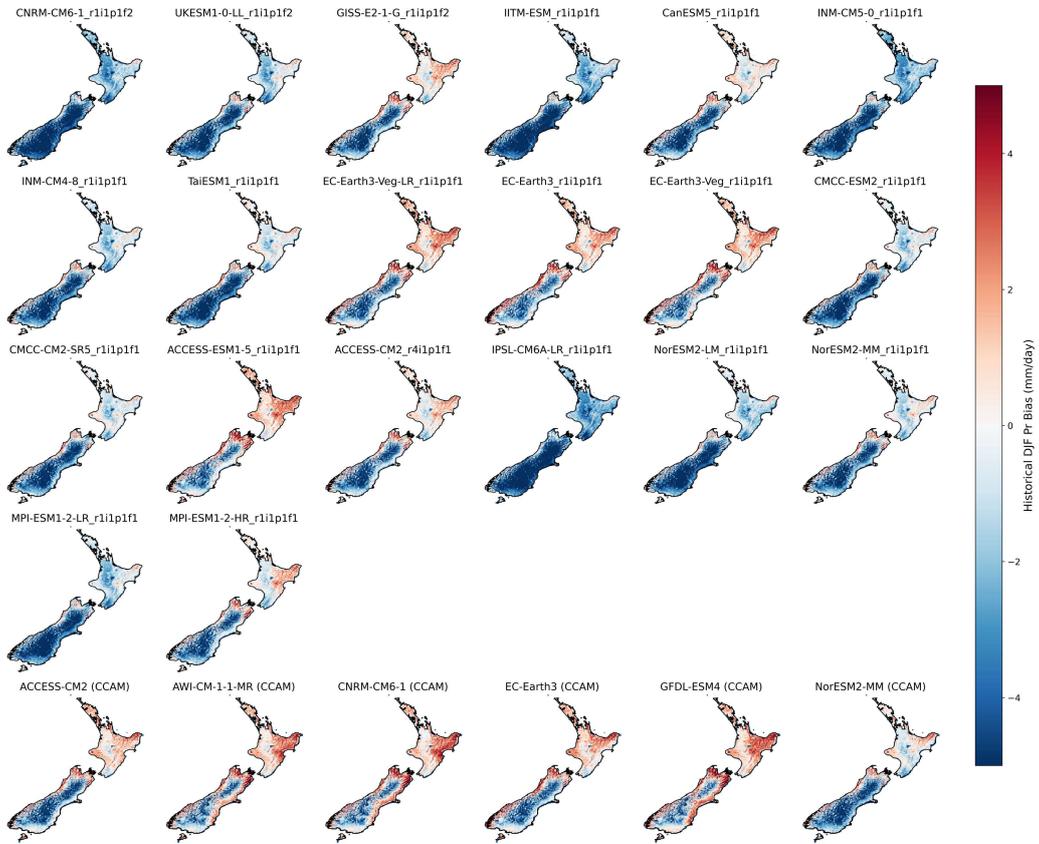

**Supplementary Figure S1**: Spatial patterns of historical climatological biases for TXx (1986-2005), comparing the emulator (top rows) and the ground truth RCM (bottom rows) relative to VCSN. Units are in °C.

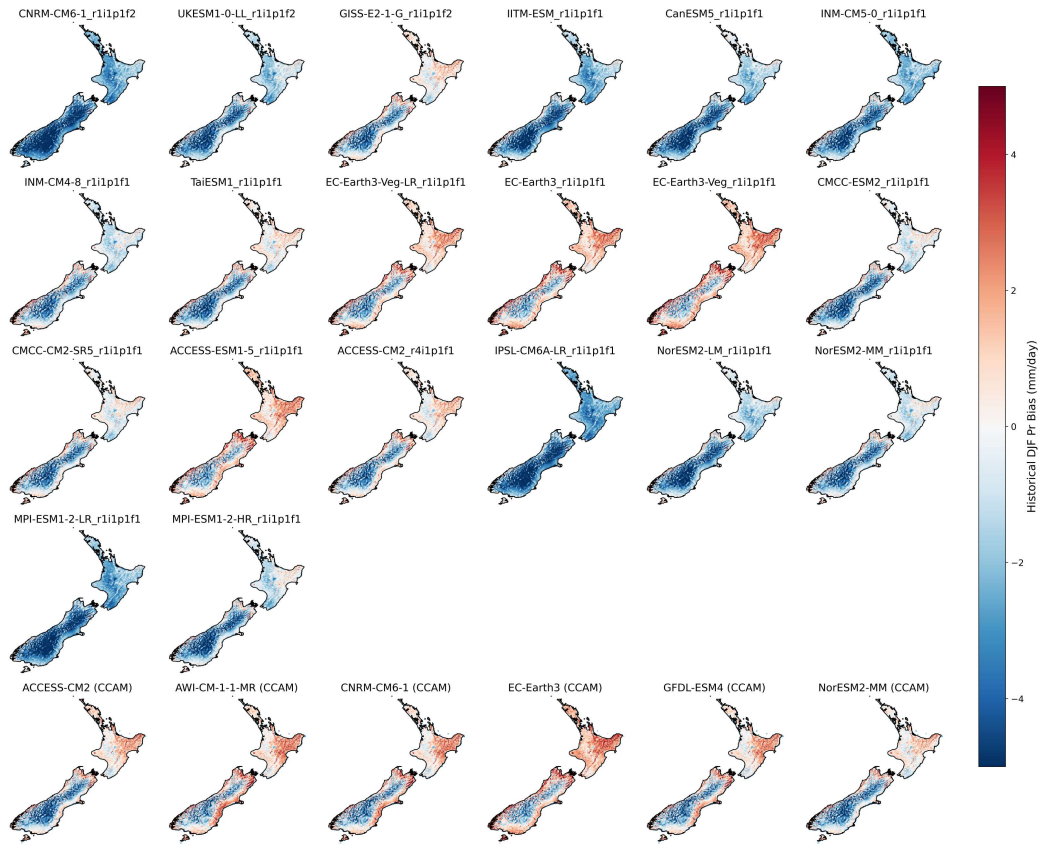

**Supplementary Figure S2**: Same as Figure S1, but for DJF Tasmax.

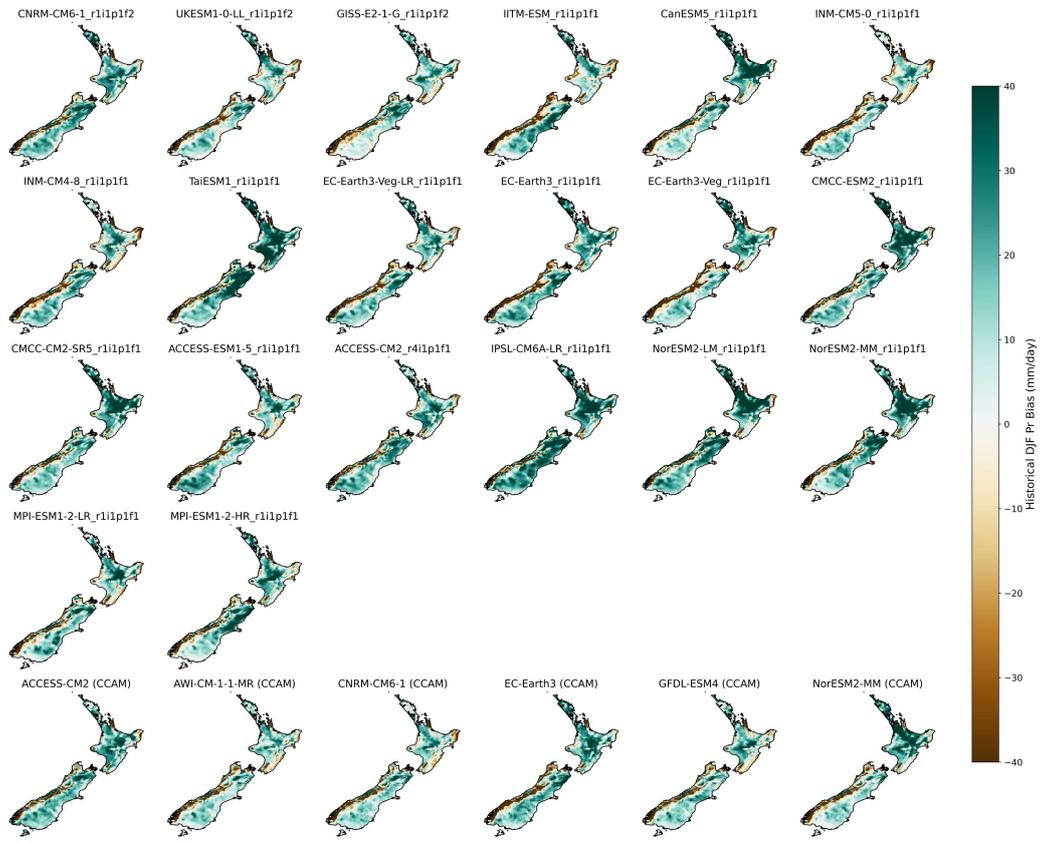

**Supplementary Figure S3**: Same as Figure S1, but for RX1Day. Errors are in units of mm/day.

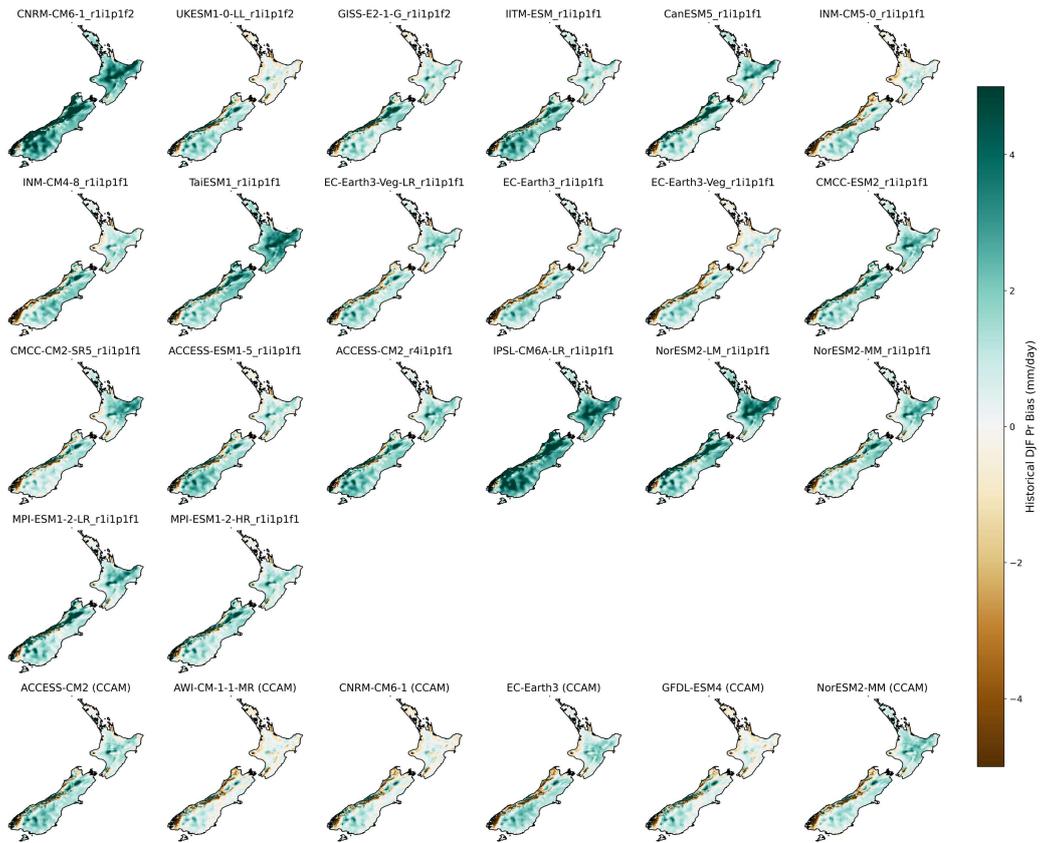

**Supplementary Figure S4**: Same as Figure S3, but for DJF seasonal mean precipitation.

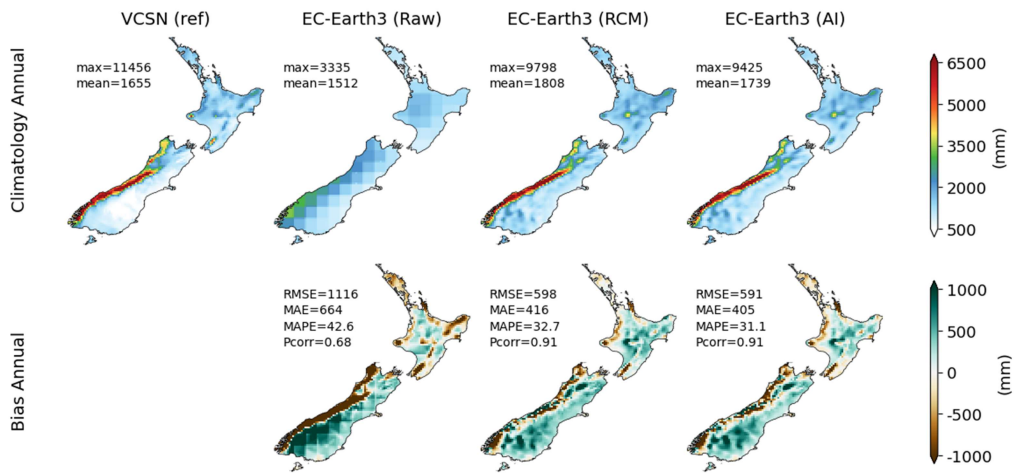

**Supplementary Figure S5**: An illustration of the improvement of downscaling (i.e. added value) for the annual mean precipitation climatology (1986-2005). The top row displays historical climatology from left to right: VCSN observations (reference), raw EC-Earth3 GCM, RCM, and the emulator (AI). The bottom row highlights biases against observations, illustrating that the emulator's (AI) biases closely resemble those of the RCM.

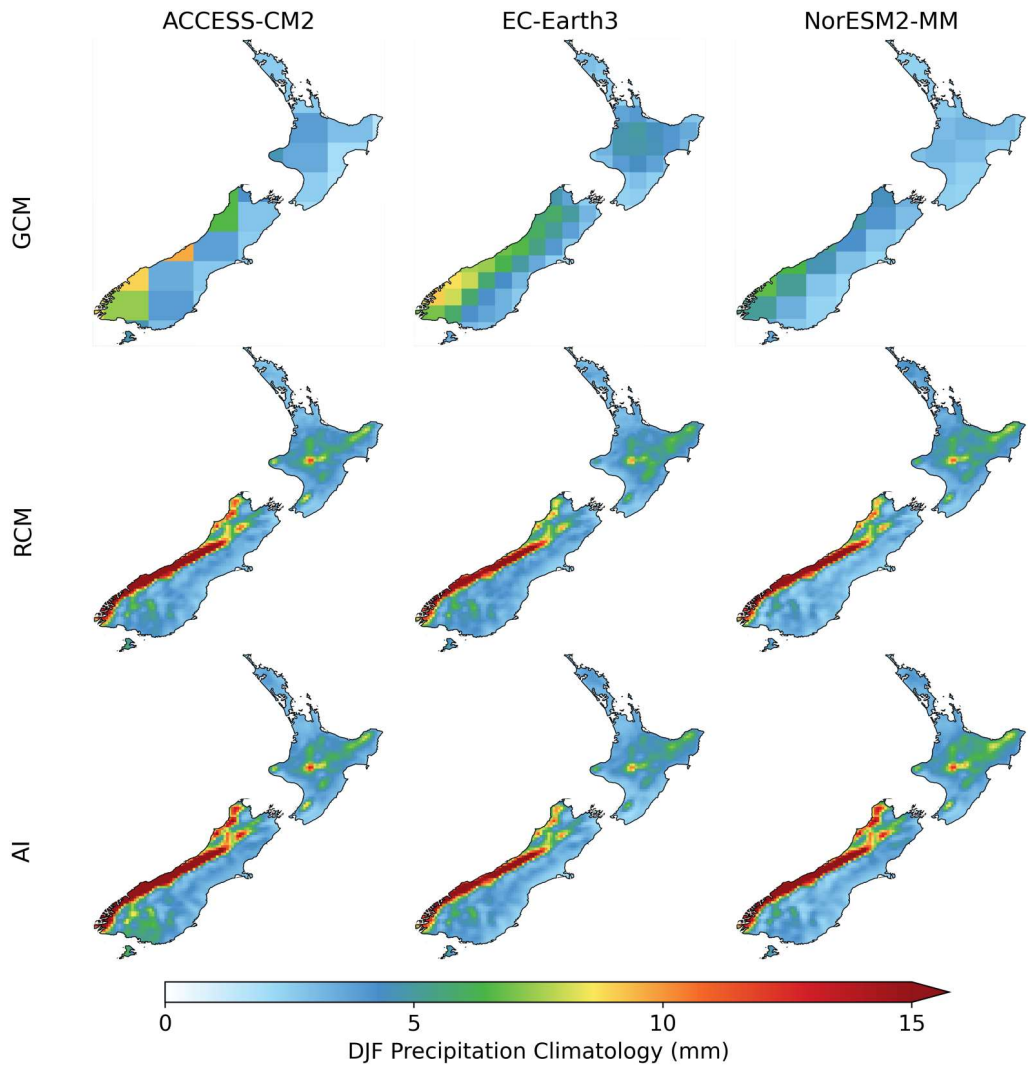

**Supplementary Figure S6**: Historical climatologies (1986-2005) of DJF precipitation climatology. The top row shows GCM fields, the second row displays the RCM, and the last row features emulator-downscaled climatologies. Note that while ACCESS-CM2 GCM was used for training, EC-Earth3 and NorESM2-MM simulations serve as out-of-sample evaluations. This figure shows that the RCM and emulator (AI) capture the climatologies with a similar level of detail.

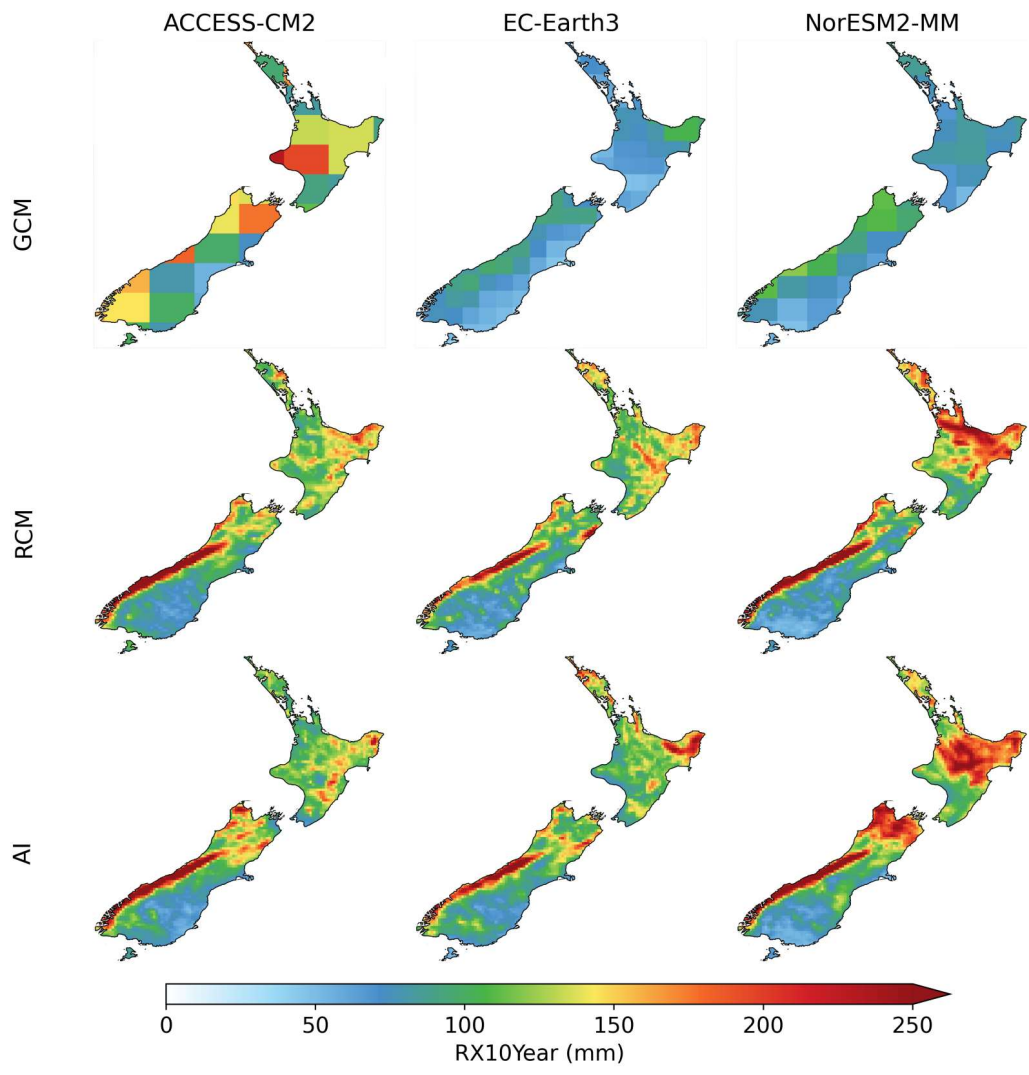

**Supplementary Figure S7**: Same as Figure S6, but for RX [10 year] (wettest-day-per-decade).

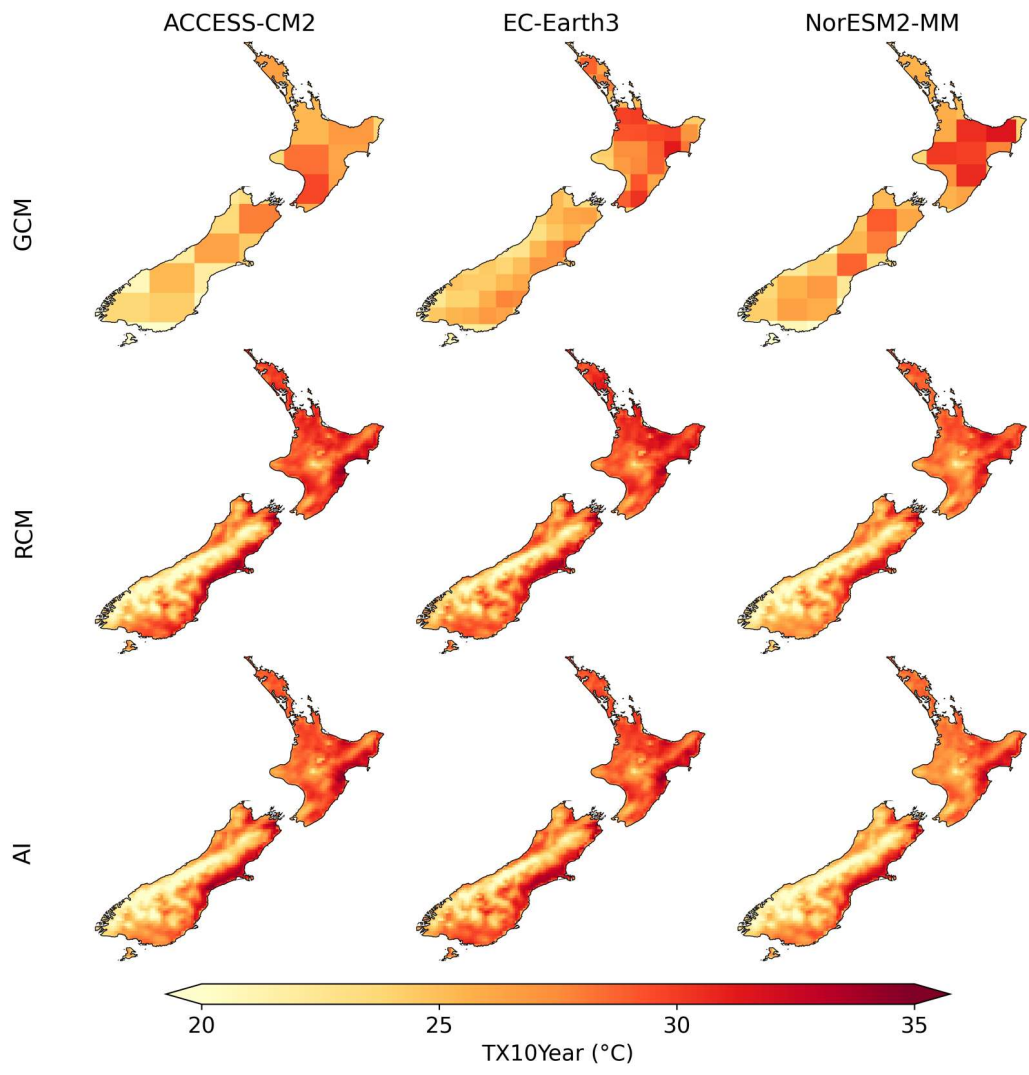

**Supplementary Figure S8**: Same as Figure S6, but for TX [10 year].

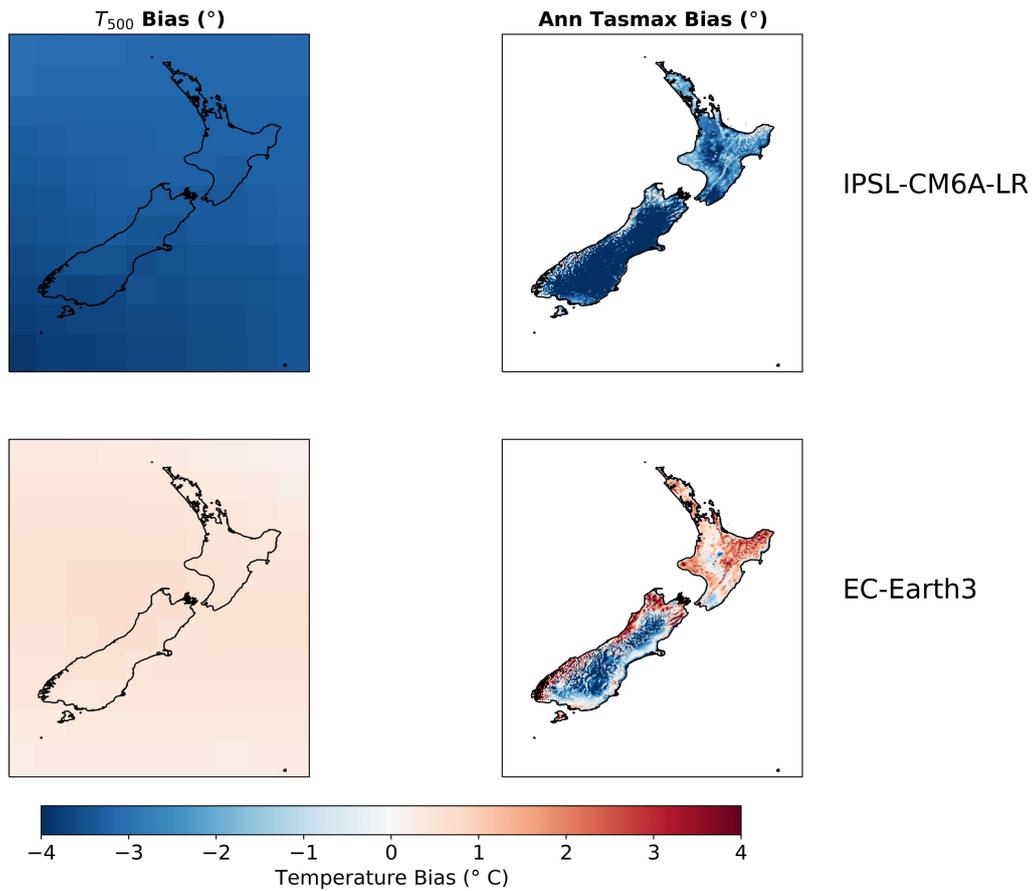

**Supplementary Figure S9**: Illustration of historical annual $T_{500}$ biases (1986–2005) in the driving GCMs (left), computed relative to ERA5 regridded to 1.5°, and how these biases are inherited by the RCM emulator during downscaling for IPSL-CM6A-LR and EC-Earth3 (right), with RCM biases computed relative to VCSN tasmax observations over the same climatology period.

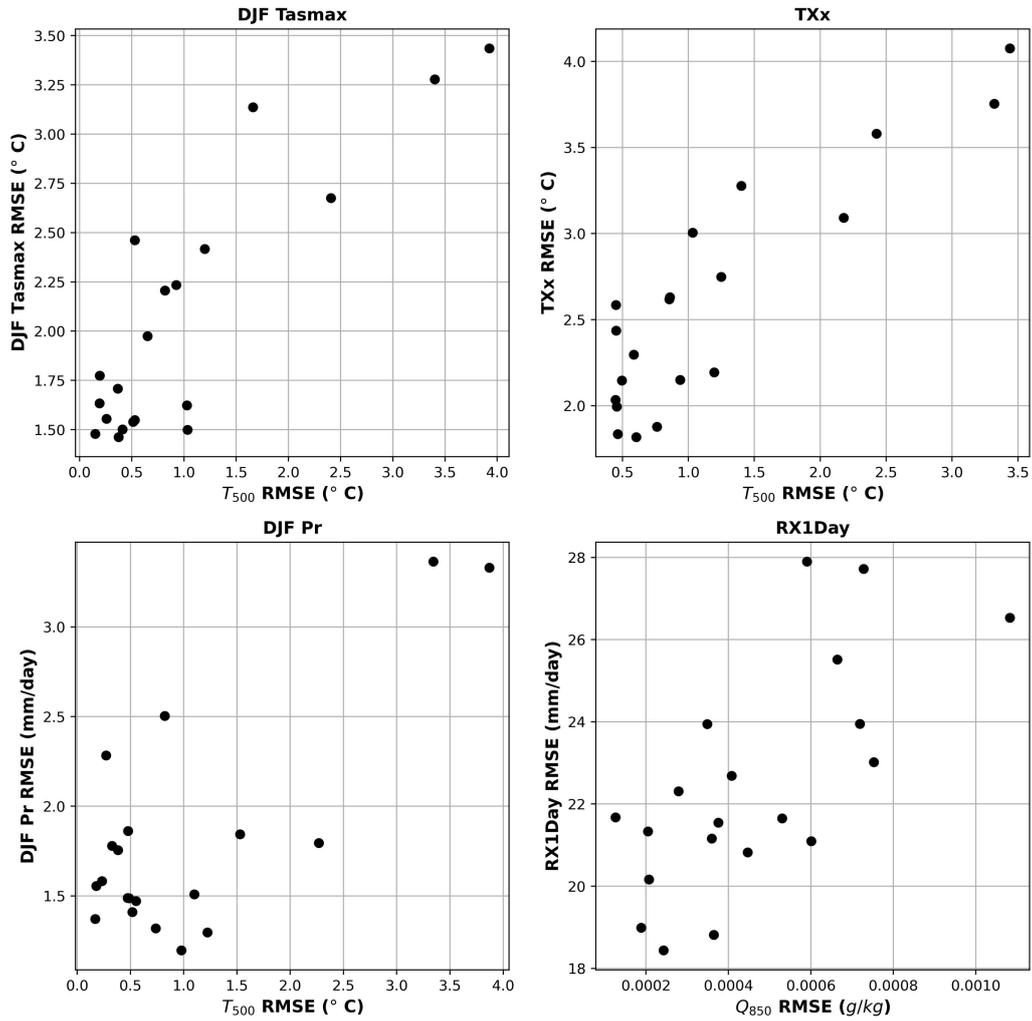

**Supplementary Figure S10**: Relationship between GCM (n=20) input biases (measured against ERA5 ground truth) and downscaled performance (measured against VCSN; observational reference), across four metrics (DJF tasmax, TXx, DJF pr, RX1Day). The x-axis displays the spatially averaged Root Mean Square Error (RMSE) of the historical annual climatology (1986-2005) for GCM inputs, relative to ERA5 ground truth (re-gridded to 1.5°×1.5°). The x-axis shows the GCM input biases for $T_{500}$, except for RX1Day uses $Q_{850}$. The y-axis shows the RMSE of downscaled outputs (DJF tasmax, TXx, DJF PR, RX1Day) relative to gridded observations. These results suggest a link between errors in GCM inputs and the performance of downscaling efforts. Correlations between the metrics and GCM inputs are shown in Figure S11.

11

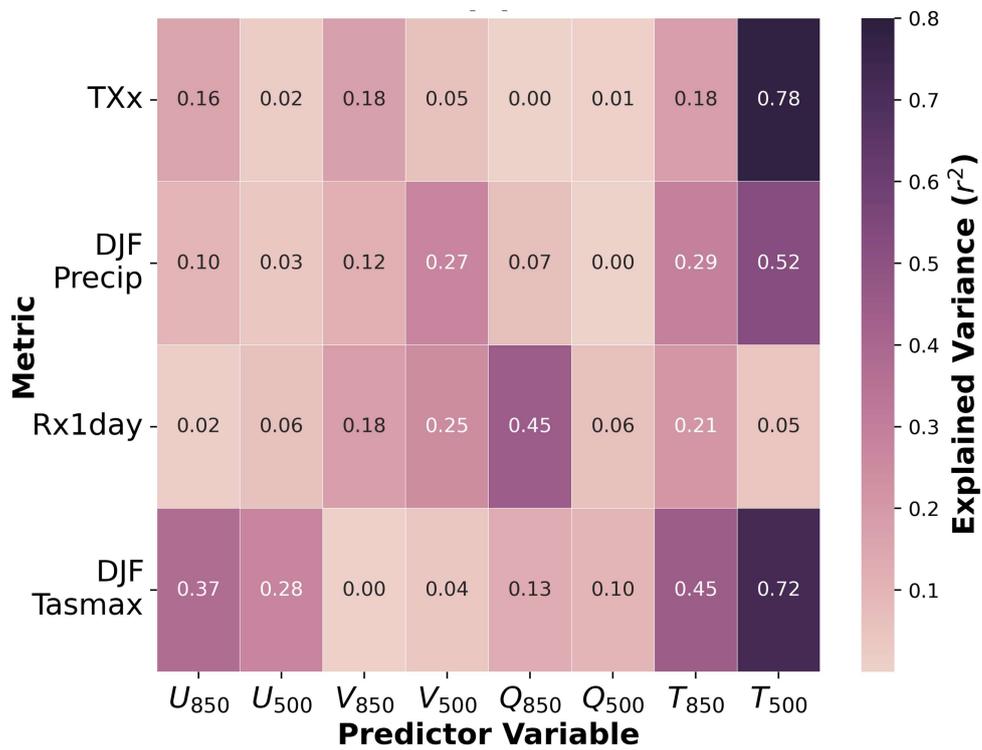

**Supplementary Figure S11**: Correlation between RMSE in GCM inputs (as in Figure S10) and downscaled biases relative to VCSN observations for each metric (y-axis). The strongest correlations are highlighted in Figure S10.

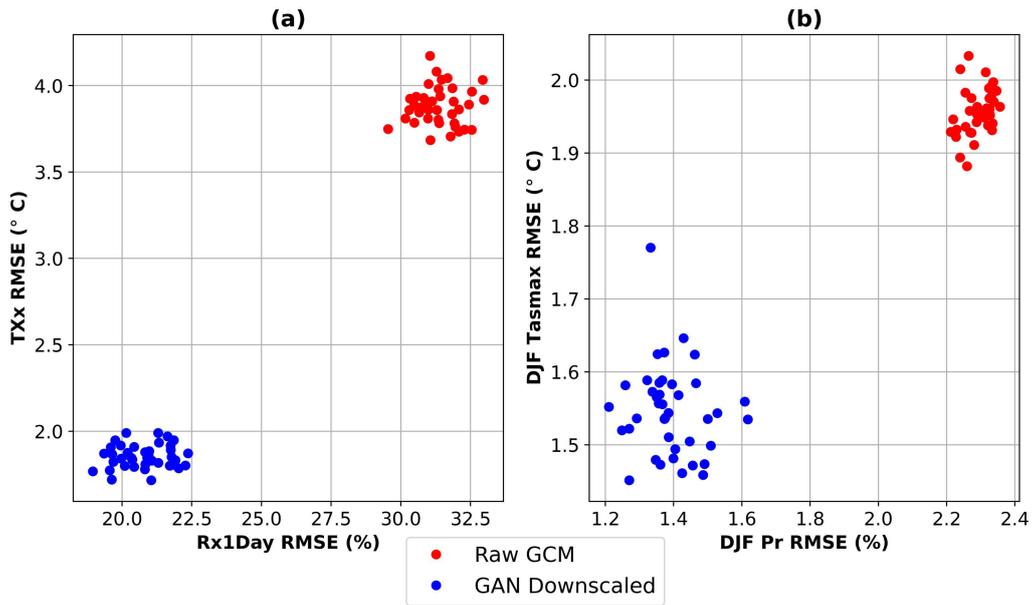

**Supplementary Figure S12**: This shows the Root Mean Square Error (RMSE) of downscaling versus no downscaling for the ACCESS-ESM1-5 ensemble (n=40). (a) TXx and RX1Day biases and (b) DJF tasmax and Pr biases are presented for both the emulator (GAN downscaled) and raw ACCESS-ESM1-5 fields. Each point denotes the bias of a single ensemble member.

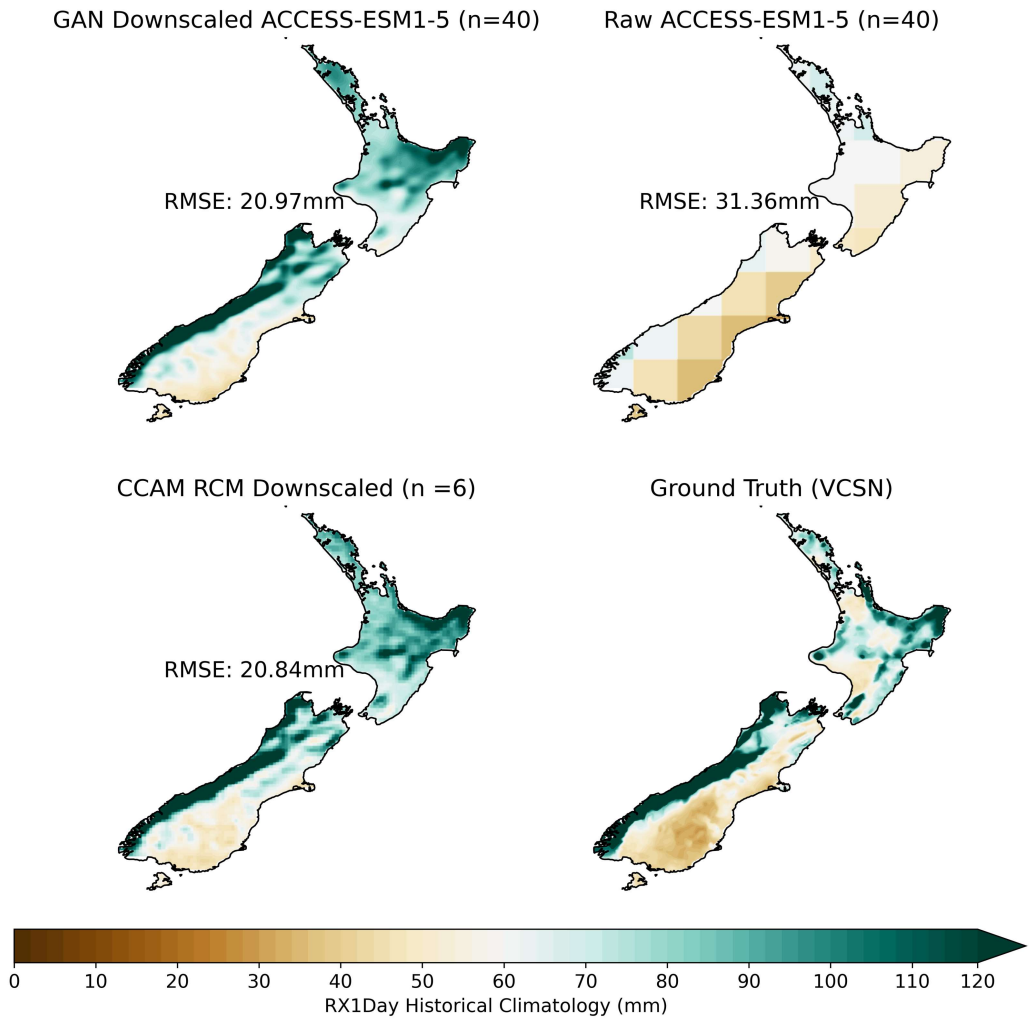

**Supplementary Figure S13**: This figure compares the historical climatology (1986-2005) of RX1Day from the ACCESS-ESM1-5 large ensemble (represented by its ensemble average) against other dynamical downscaled simulations and observations. It compares the GAN downscaled ensemble (top left), raw ACCESS-ESM1-5 (top right), and CCAM (bottom left) against the observed climatology (bottom left, ground truth). RMSE values are computed relative to the ground truth. CCAM RCM downscaled simulations (an average of six historical GCMs, not ACCESS-ESM1-5) provide a general benchmark for RCM skill.

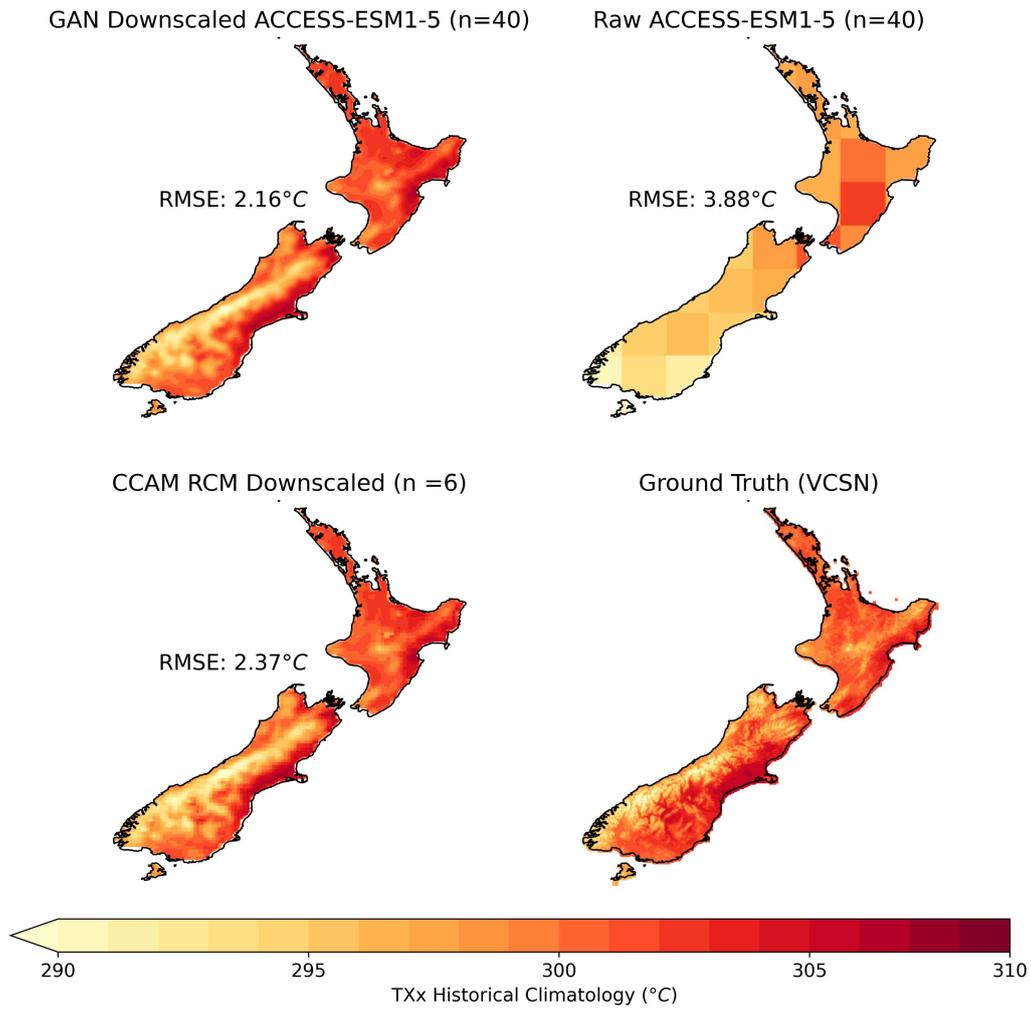

**Supplementary Figure S14**: Same as S13, but for TXx.

## Climate Change Signal Evaluation (Figures S15-S22)

This section evaluates the emulator's skill in capturing climate changes—defined as differences between future and historical climates—for seasonal means, annual extremes, and decadal extremes.

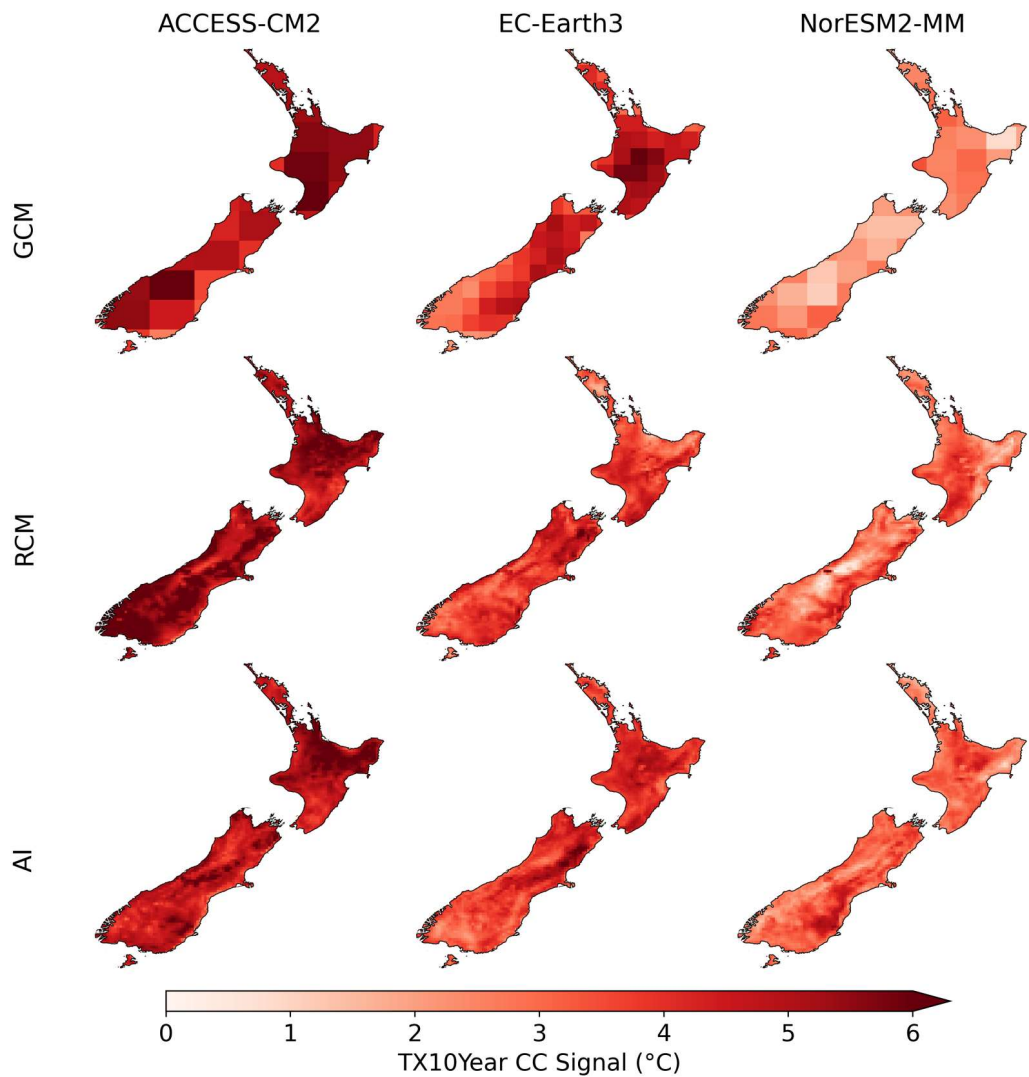

**Supplementary Figure S15**: Spatial patterns of the climate change signal in TX [10 year], defined as the difference between future (2080–2099) and historical (1986–2005) values under SSP3-7.0. The top row shows GCM outputs, the second row shows RCM results, and the third row shows emulator-downscaled signals. Results indicate that the emulator closely matches the RCMs for two out-of-sample GCMs (EC-Earth3 and NorESM2-MM), while ACCESS-CM2 was included in training.

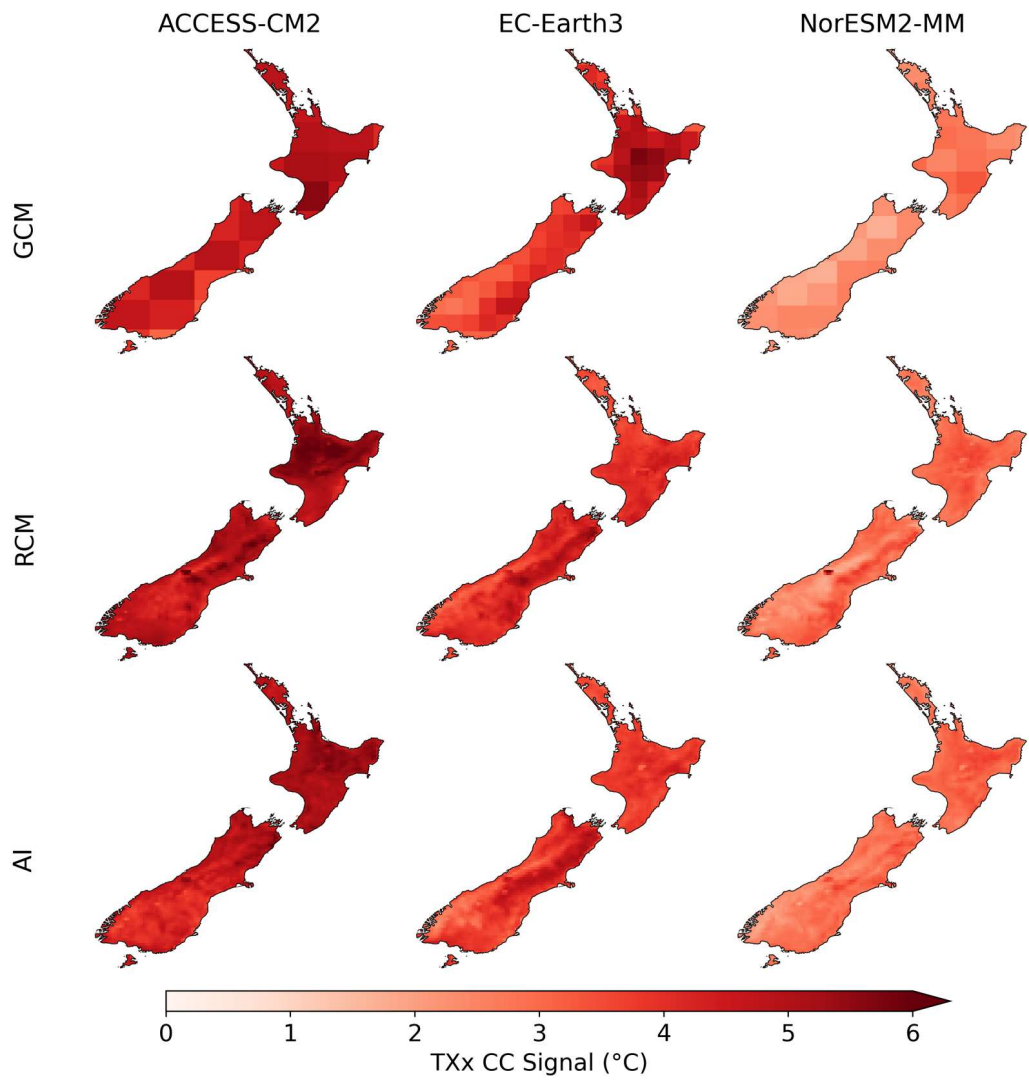

**Supplementary Figure S16**: Same as Figure S15, but for TXx.

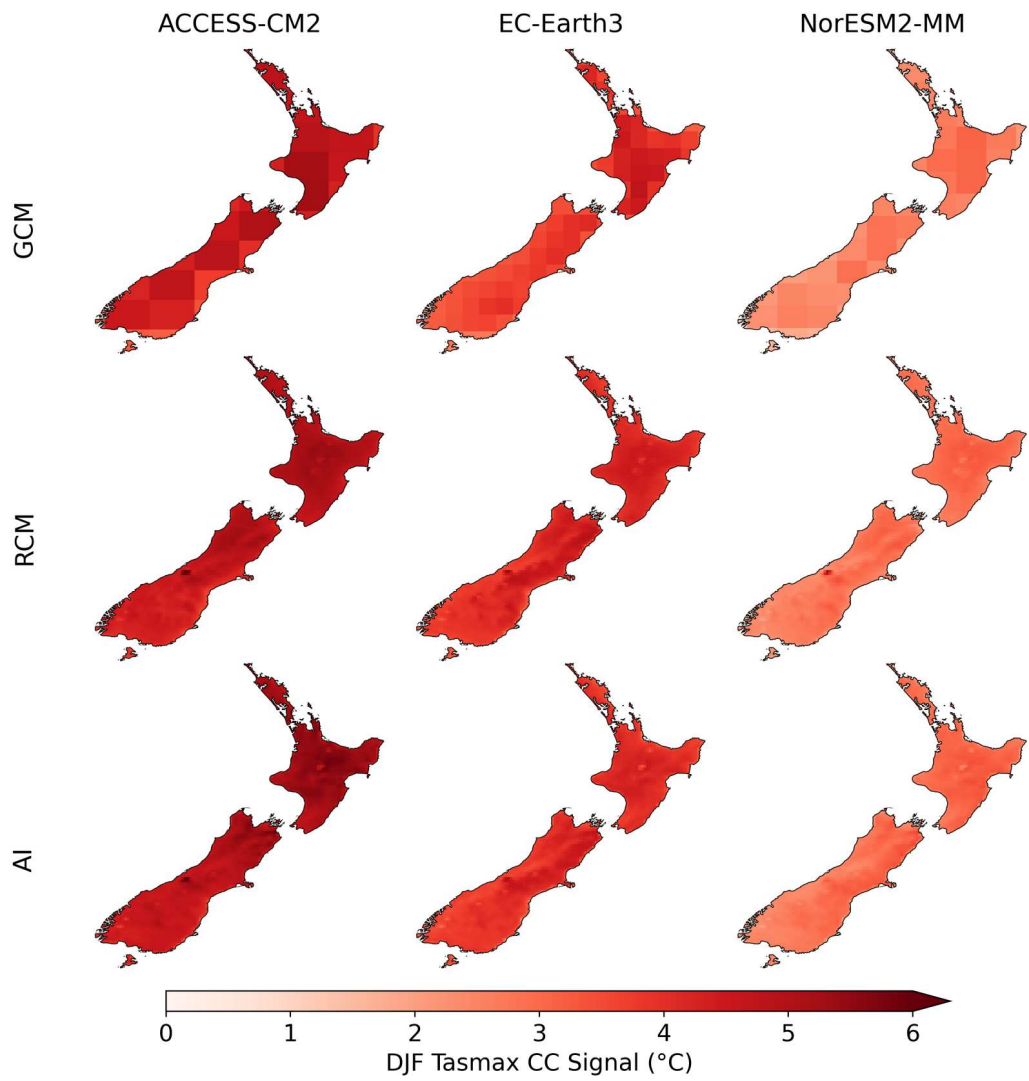

**Supplementary Figure S17**: Same as Figure S15, but for DJF Tasmax.

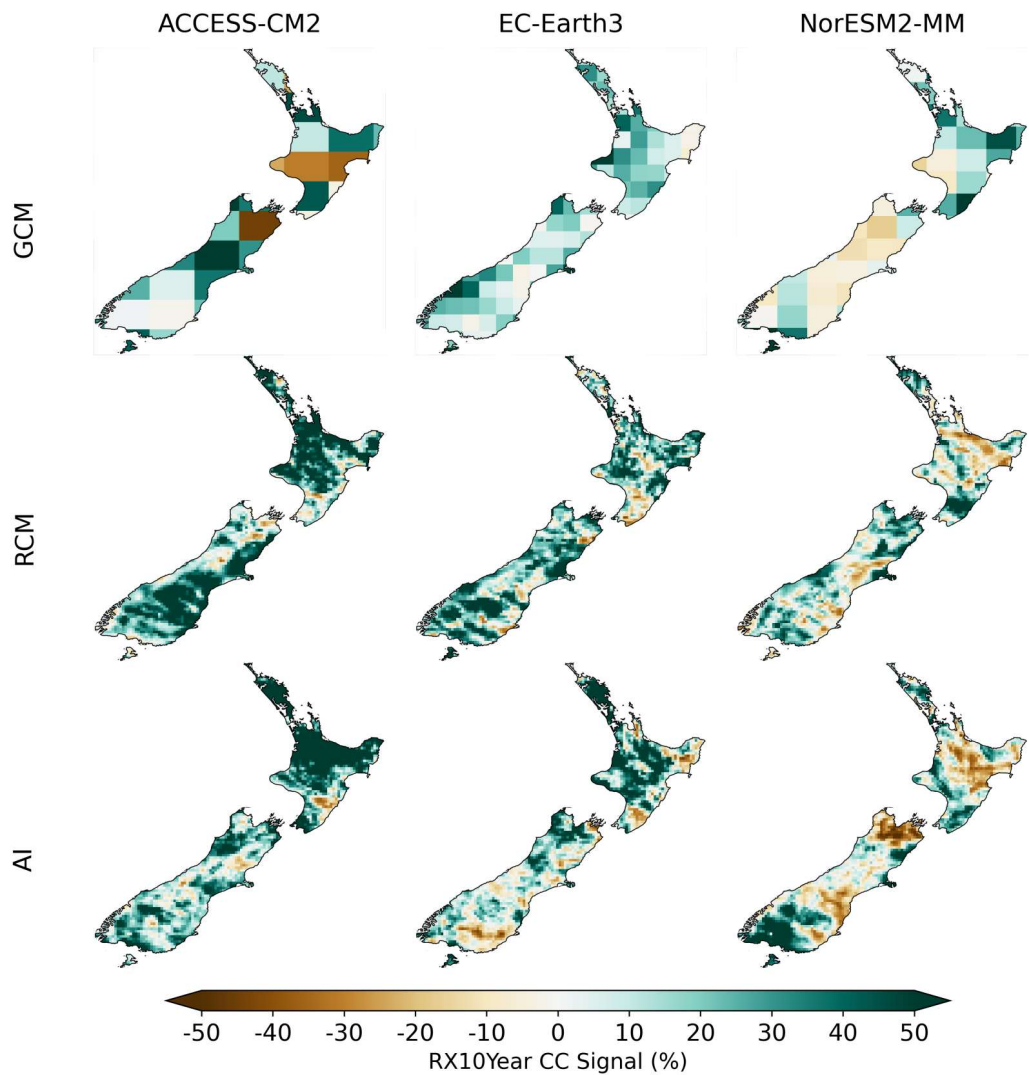

**Supplementary Figure S18**: Same as Figure S15, but for RX [10 year], and the climate change signal is the % difference between future and historical.

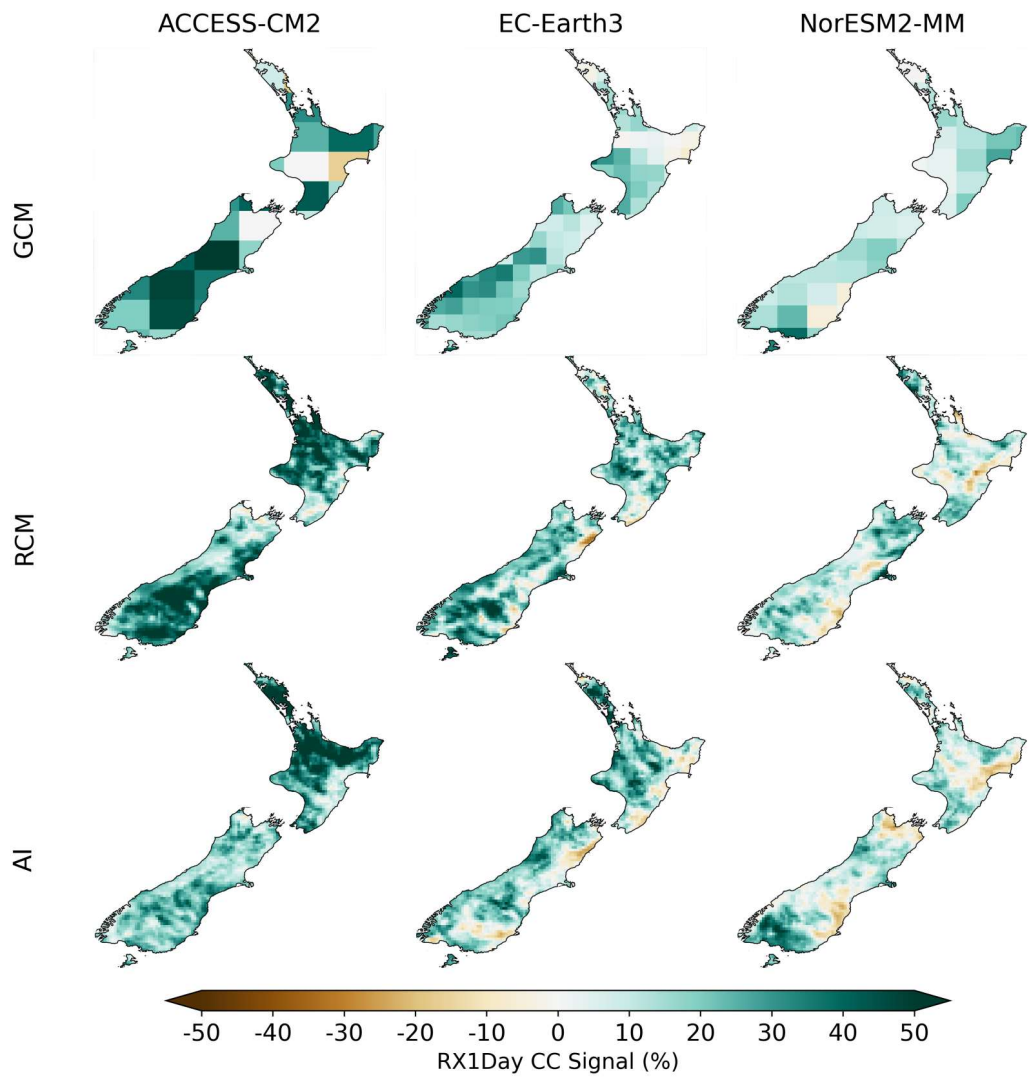

**Supplementary Figure S19**: Same as Figure S18, but for RX1Day.

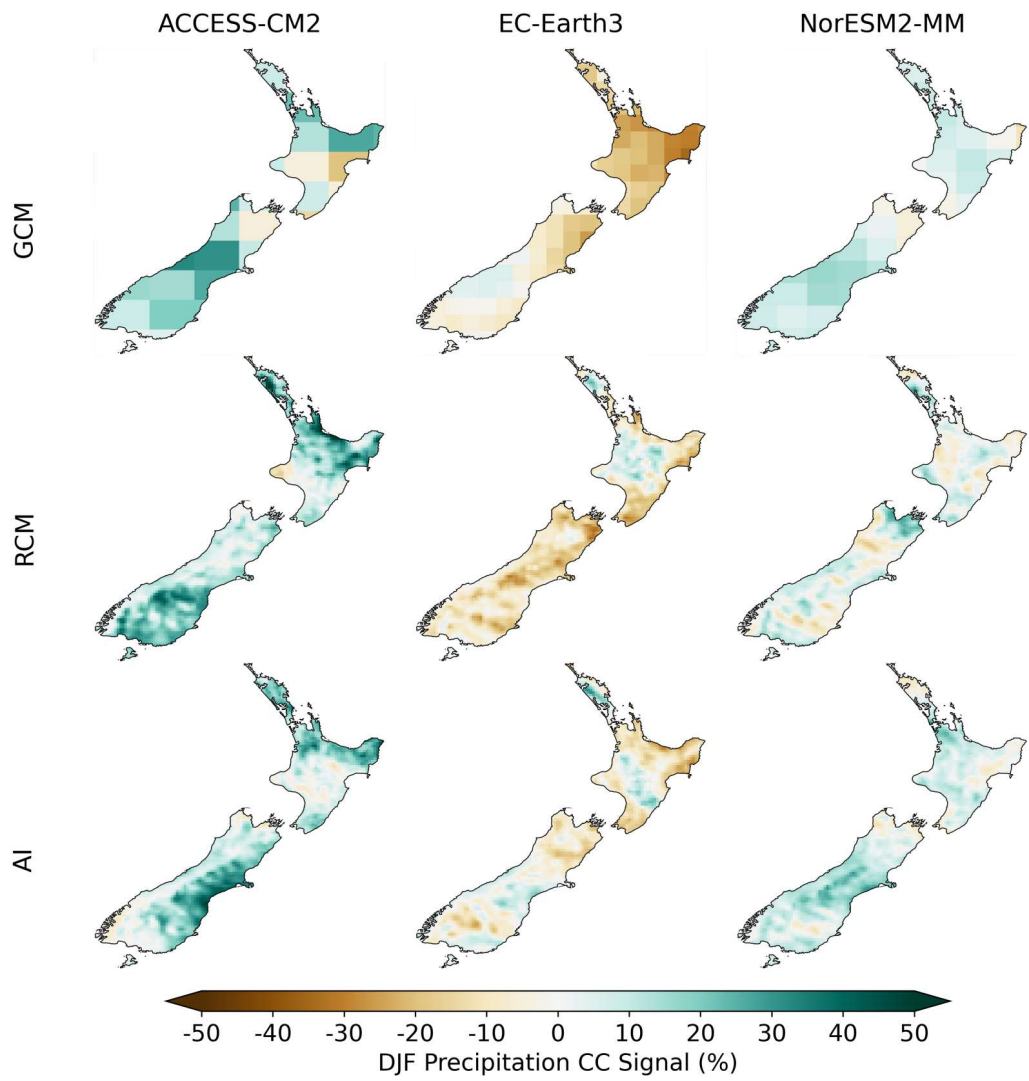

**Supplementary Figure S20**: Same as Figure S18, but for DJF seasonal mean precipitation.

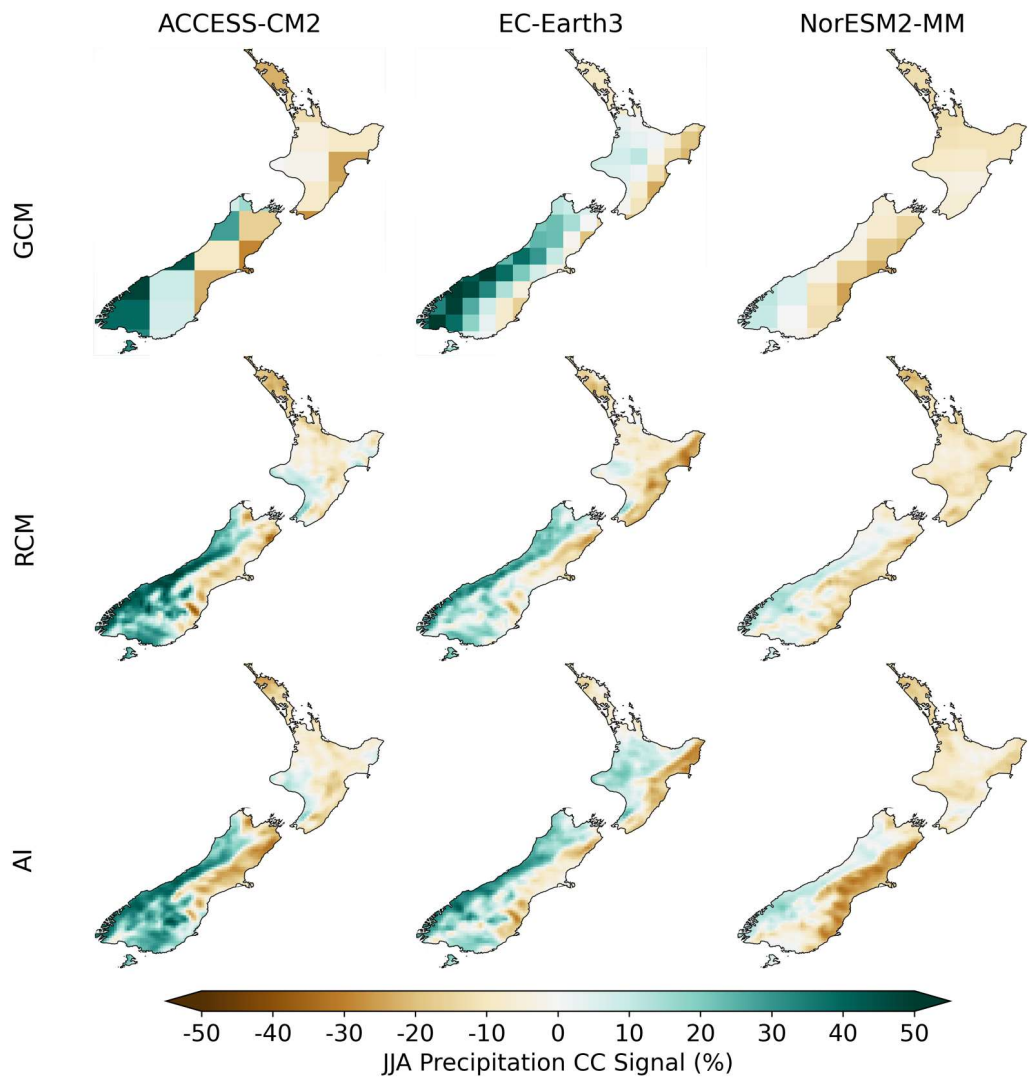

**Supplementary Figure S21**: Same as Figure S18, but for JJA seasonal mean precipitation.

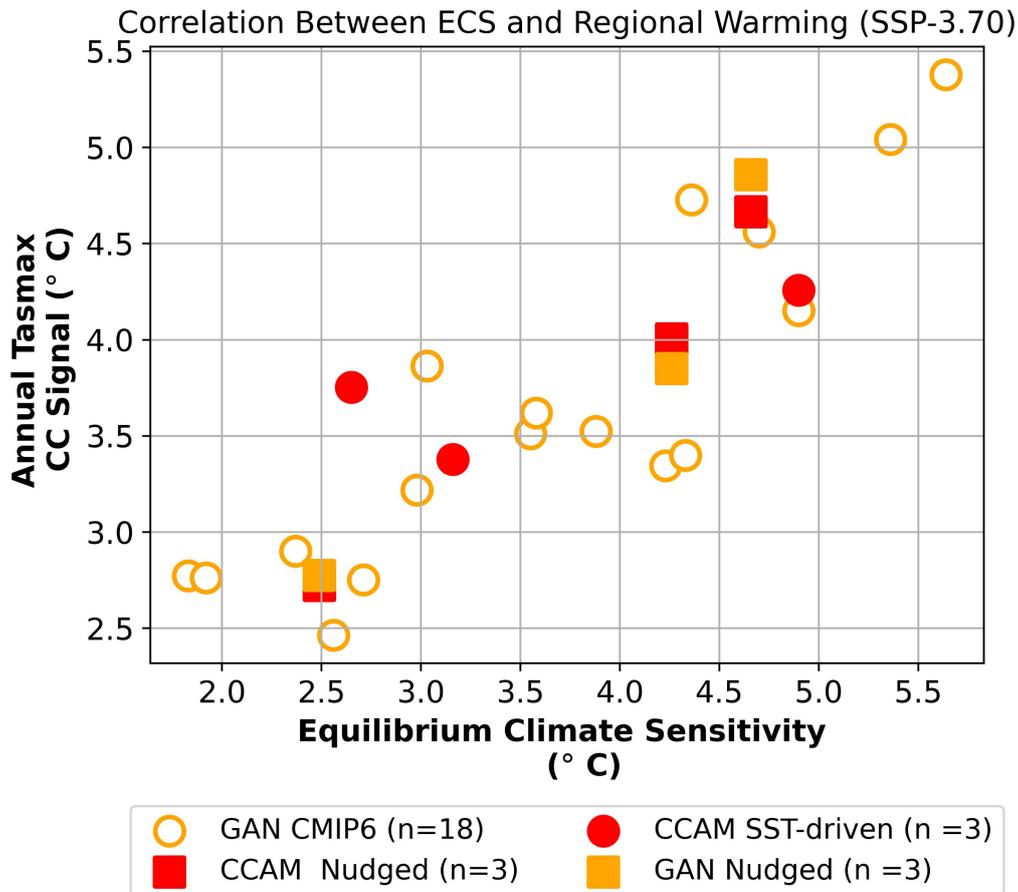

**Supplementary Figure S22**: Illustration of the relationship between GCM Equilibrium Climate Sensitivity (x-axis) and the emulator-downscaled annual temperature change (2080–2099 vs. 1986–2005) for the SSP3-7.0 scenario, averaged over land. Red squares show the three CCAM-nudged RCM simulations (ACCESS-CM2, EC-Earth3, NorESM2-MM), with corresponding GAN-downscaled signals (orange squares). Red circles represent SST-driven simulations not directly comparable to GAN outputs. Unfilled orange circles for the remaining 17 CMIP6 GCMs.

## A range of possible futures for temperature and precipitation extremes (Figures S23-S27)

This section shows ensemble variance in the ACCESS-ESM1-5 and CanESM5 ensembles to complement the section "A range of possible futures for temperature and precipitation extremes".

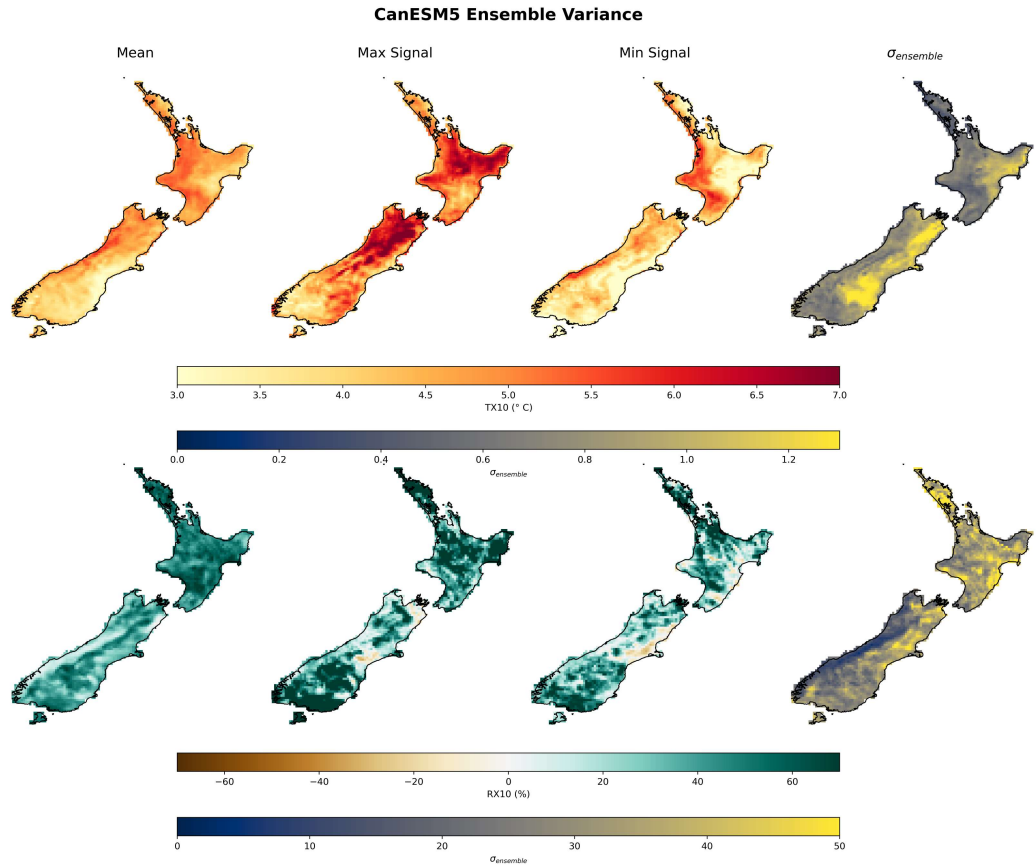

**Supplementary Figure S23**: Same as Figure 2 in the main text, but for the CanESM5 ensemble, showing only the emulator-downscaled outputs at 12 km resolution.

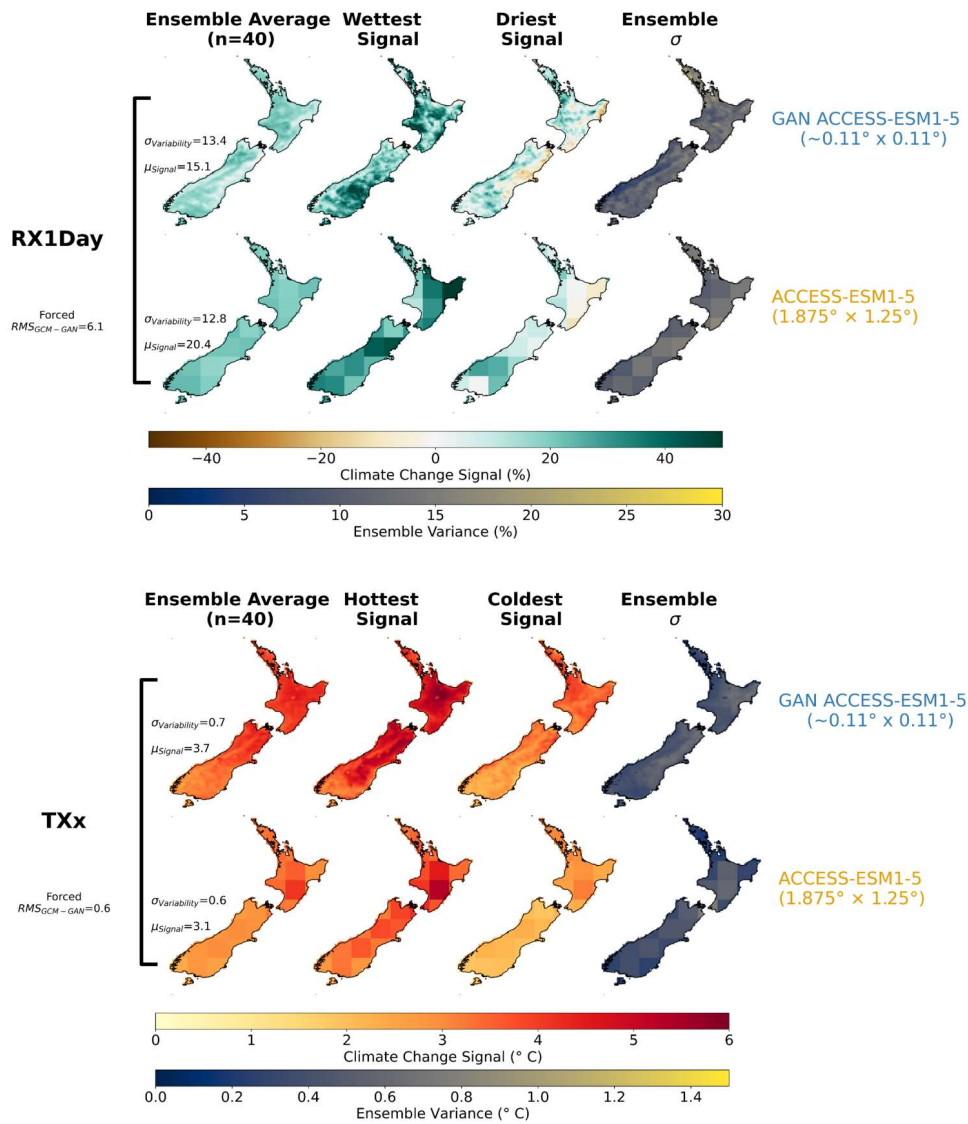

Supplementary Figure S24: Same as Figure 2 in the main text, but for annual extremes (RX1Day, TXx) in the ACCESS-ESM1-5 ensemble.

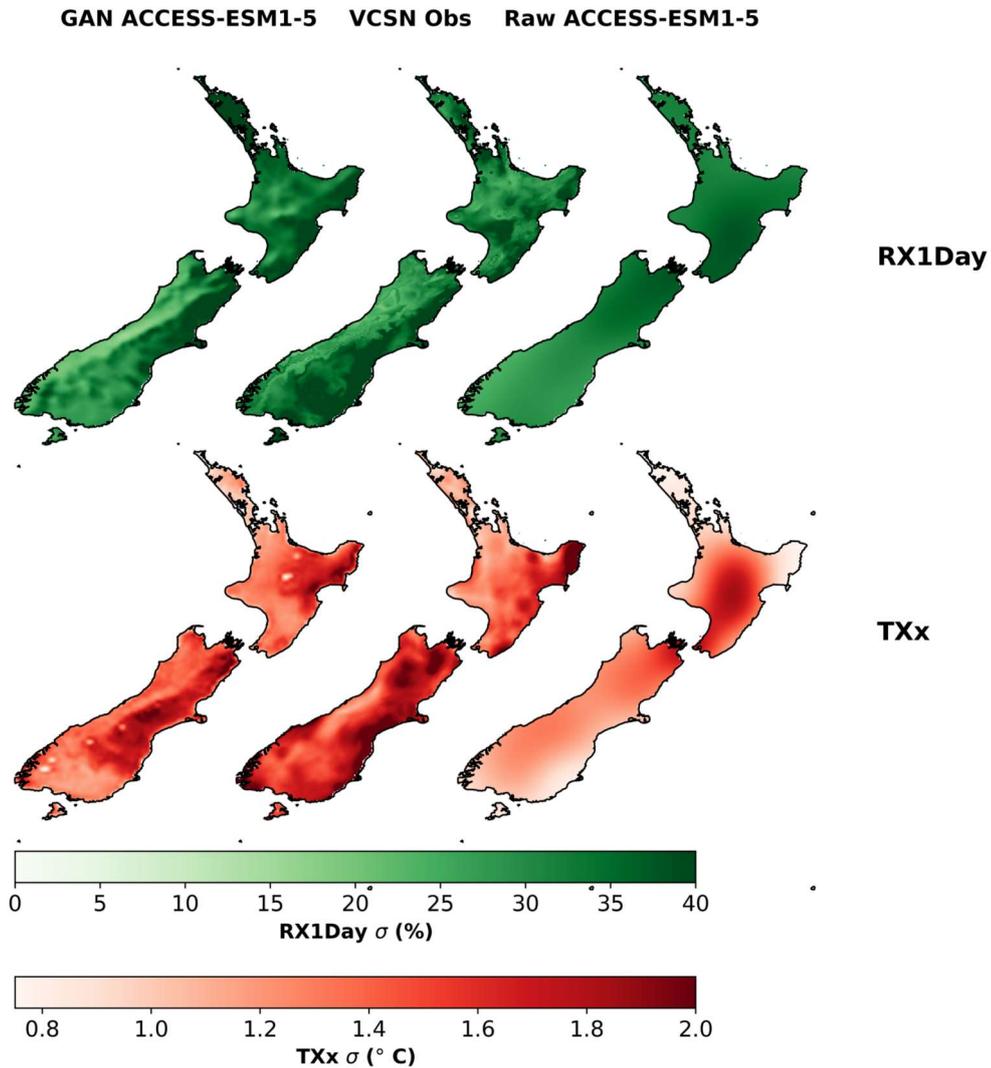

**Supplementary Figure S25**: Historical interannual variability (standard deviation over 1973–2014) in downscaled annual extremes from ACCESS-ESM1-5 (averaged across all members), compared with VCSN observations and raw ACCESS-ESM1-5. Variability is computed after removing the mean over this period, using absolute values for temperature and percentage anomalies for precipitation. This Figure is directly comparable to the ensemble variance in Figure 2.

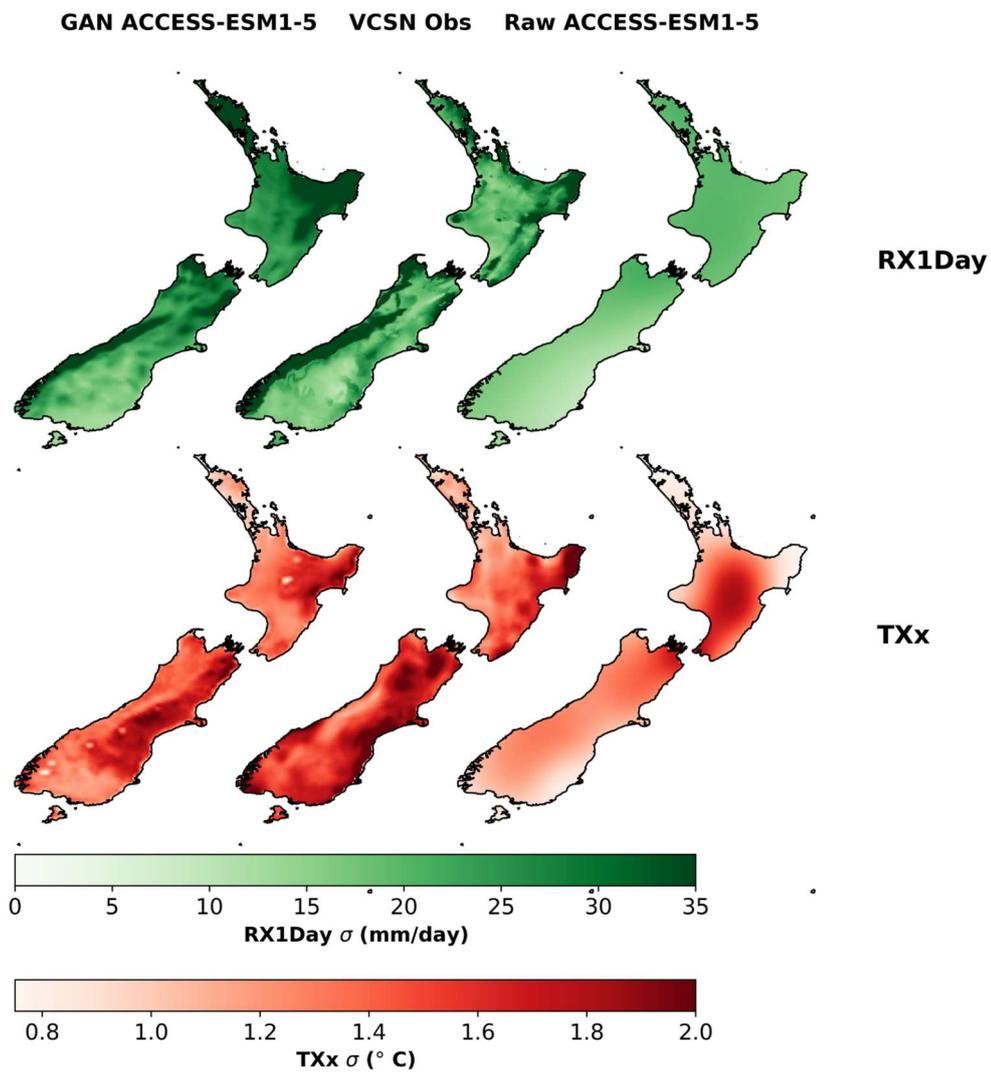

**Supplementary Figure S26**: Same as Figure S25 but using absolute anomalies for RX1Day as a reference.

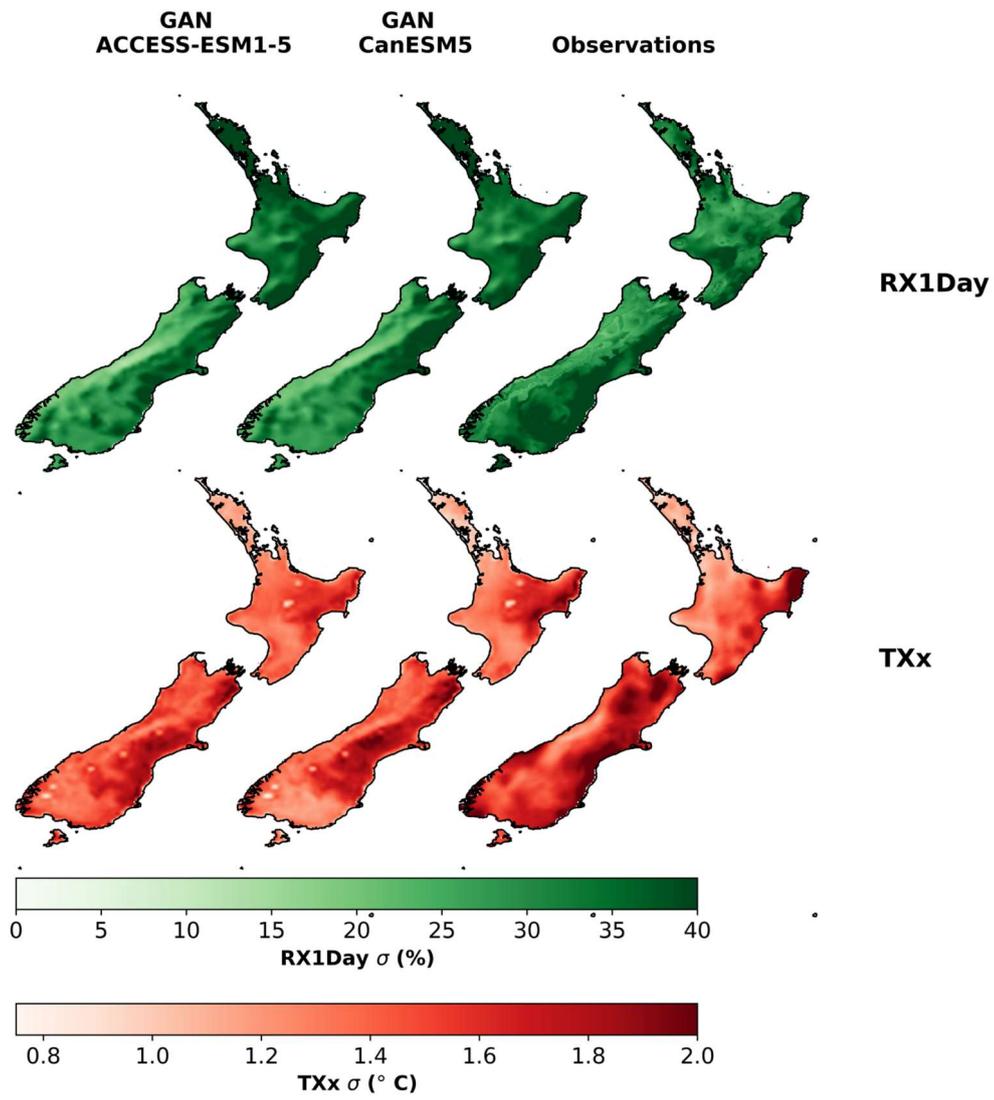

**Supplementary Figure S27**: Same as Figure S25 but including CanESM5.

# Uncertainty decomposition at fine scales (Figures S28-S31)

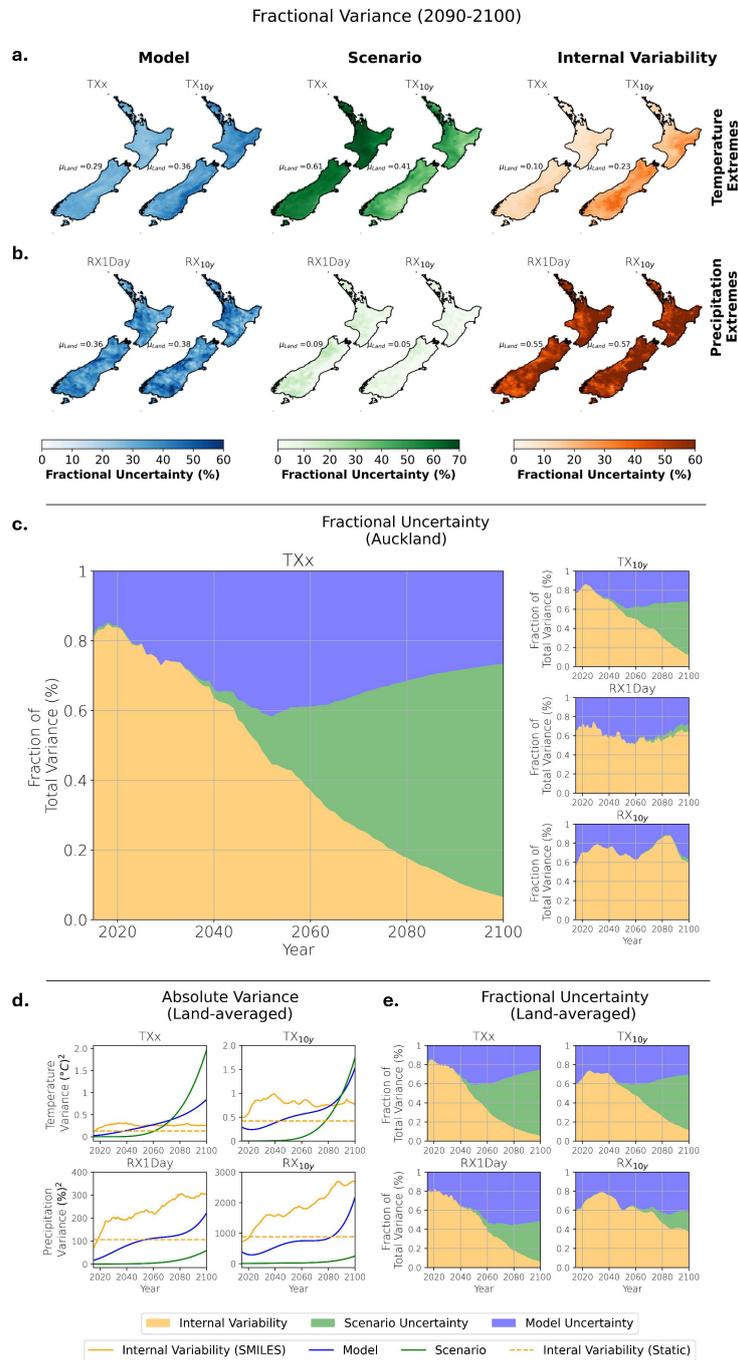

**Supplementary Figure S28**: Same as in Fig 4. in the main text, but using percentage anomalies instead

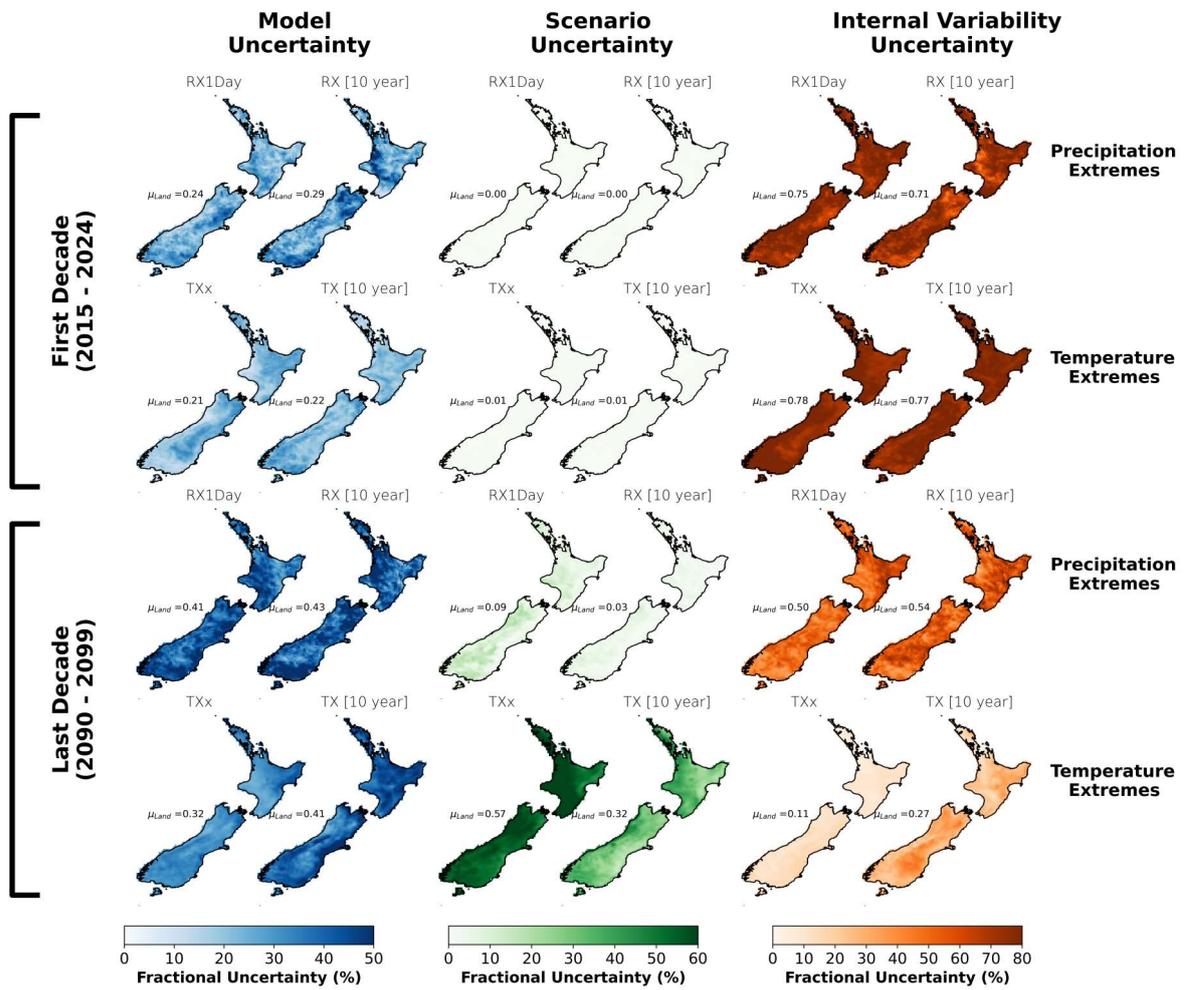

**Supplementary Figure S29**: Same as in Figure 4 in the main text (top row), but for the first decade of the projection period (2015-2024).

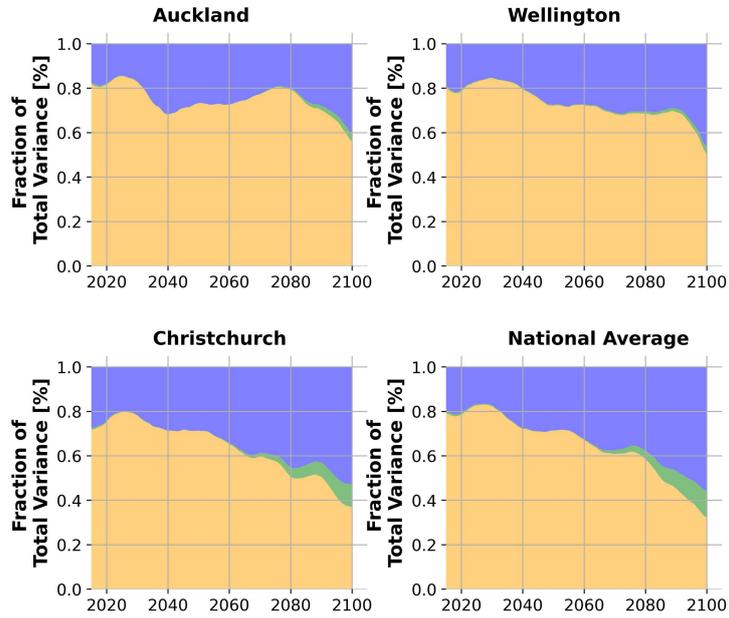

(a)

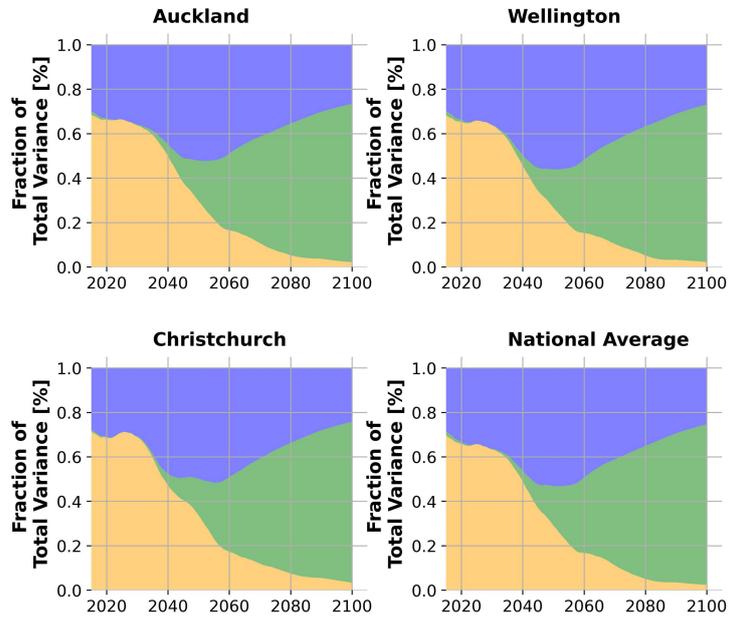

(b)

**Supplementary Figure S30**: Same as Figure S28 (percentage anomalies), but for seasonal mean DJF precipitation (top two rows) and tasmax (bottom two rows) for individual cities across New Zealand. As described in the main text, national average fractional uncertainty is calculated by first averaging trends across all land grid points before applying the fractional decomposition

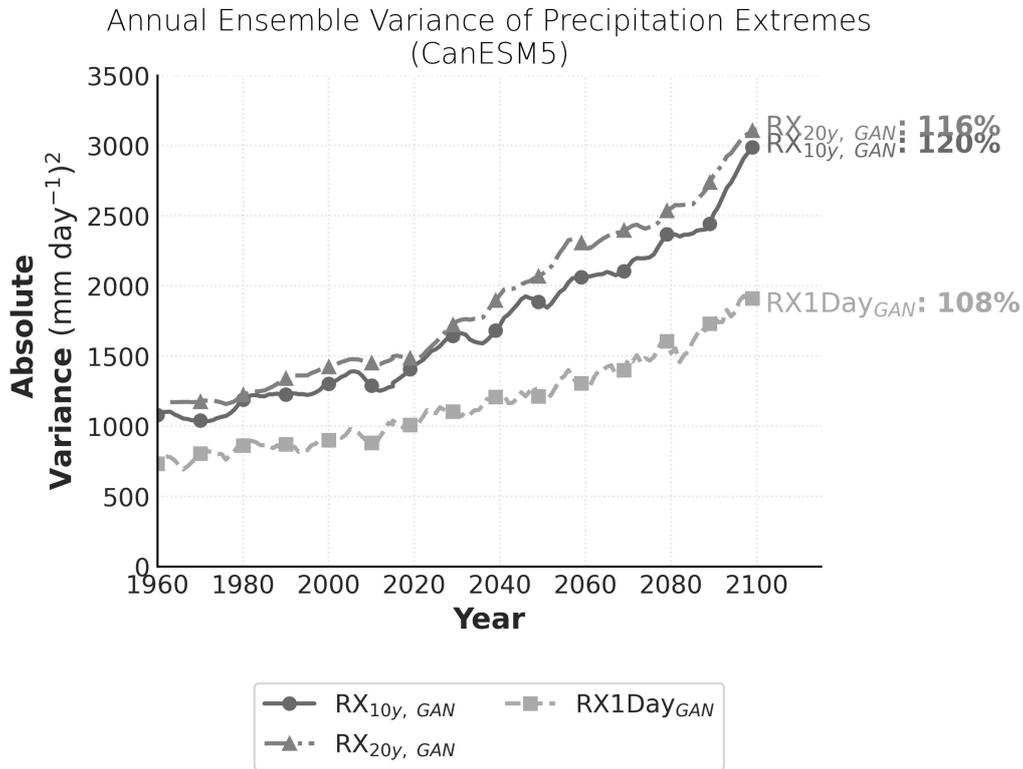

Supplementary Figure 31: Same as in Figure 5 but for CanESM5

# Online Methods

## A comparison of the perfect vs imperfect framework (Figures S32-S35)

This section provides supplementary material for the Online Methods section. We've previously shown that our emulator generates plausible climate change signals consistent with RCMs. Here, we aim to demonstrate how biases change when the emulator is applied to coarsened-RCM fields (perfect application) versus directly to GCM inputs (imperfect application). We also assess if these differences impact historical climatology and the emulator's ability to produce climate change signals. We evaluate the emulator's out-of-sample performance using three GCMs: ACCESS-CM2 (used in training), and the unseen EC-Earth3 and NorESM2-MM, for which ground truth RCM simulations are available. Figures S32-S35 illustrate the differences between perfect and imperfect biases relative to ground truth RCMs across these three GCMs. This comparison covers annual maximum surface air temperature (tasmax), precipitation, and extreme indices like RX1Day (maximum 1-day precipitation) and TXx (annual maximum daily maximum temperature). Generally, biases are slightly

higher in the imperfect framework, as indicated by the Root Mean Square Error (RMSE) values on the plots. However, signal errors are very similar. This suggests that while subtle mismatches in GCM inputs can affect the circulation (which RCMs are known to modify),

and, consequently, temperature and precipitation climatology, the climate change signal itself is well-preserved. We observe broadly similar bias patterns in the emulator, though they do vary between GCMs.

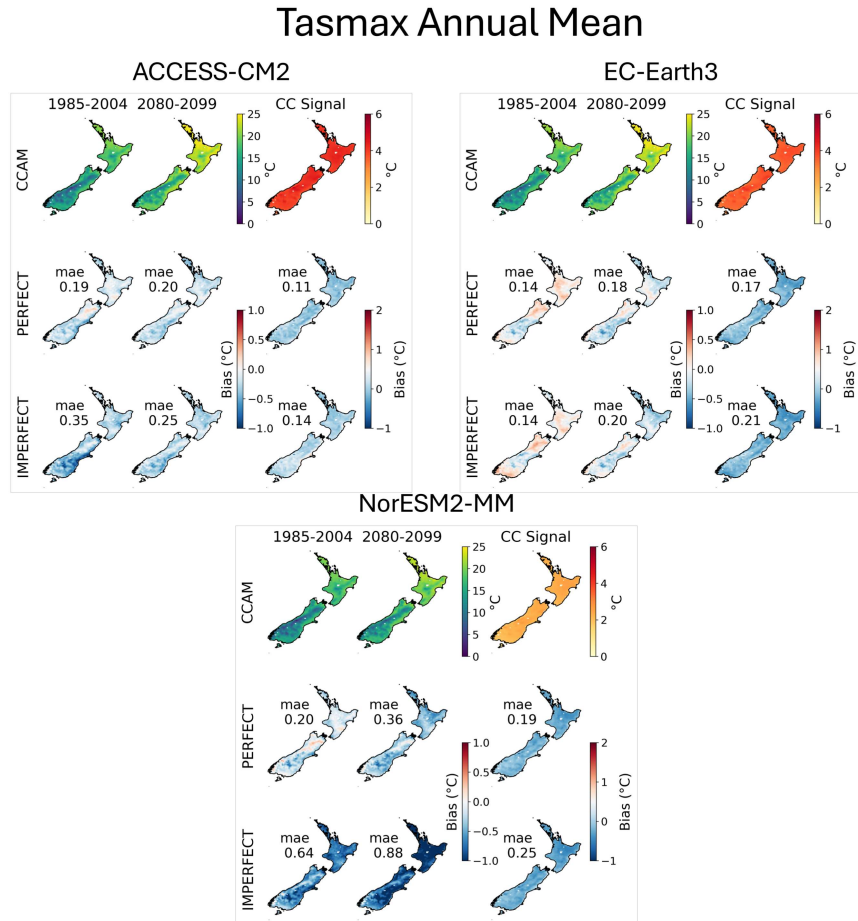

**Supplementary Figure S32**: Comparison of biases (difference between emulator and ground truth) in the historical climatology (1985–2004), future climatology (2080–2099), and the climate change signal (future minus historical) for three GCMs: ACCESS-CM2, EC-Earth3, and NorESM2-MM. Each plot shows: (top) the RCM ground truth (CCAM), (middle) emulator biases in the perfect framework (relative to CCAM), and (bottom) biases in the imperfect framework (emulator applied directly to GCM fields). The RMSE is shown in text in each of the subplots.

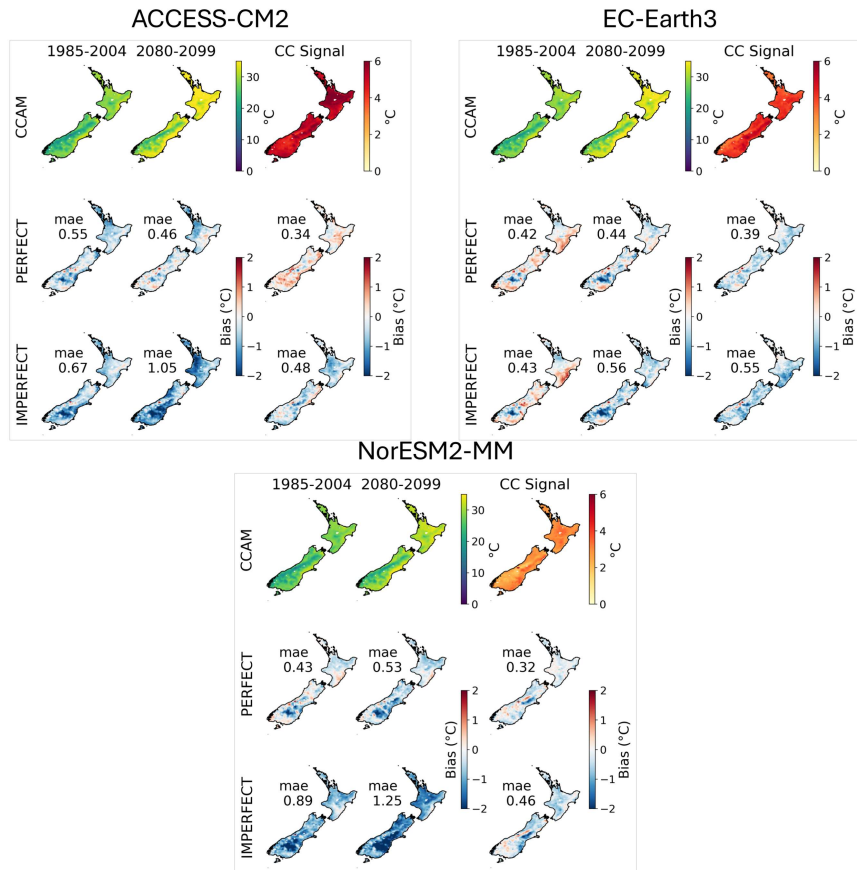

**Supplementary Figure S33**: Same as Figure S32 but for TXx (hottest day of the year).

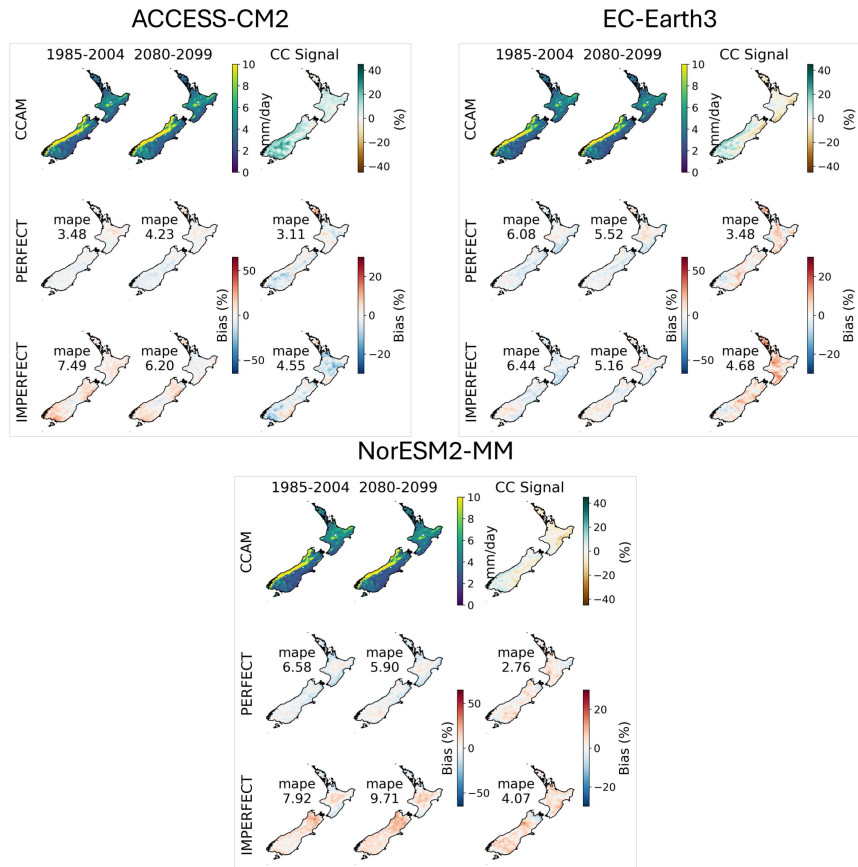

**Supplementary Figure S34**: Same as Figure S32, but for annual mean precipitation. Note, the climate change signal is computed as a % difference for precipitation.

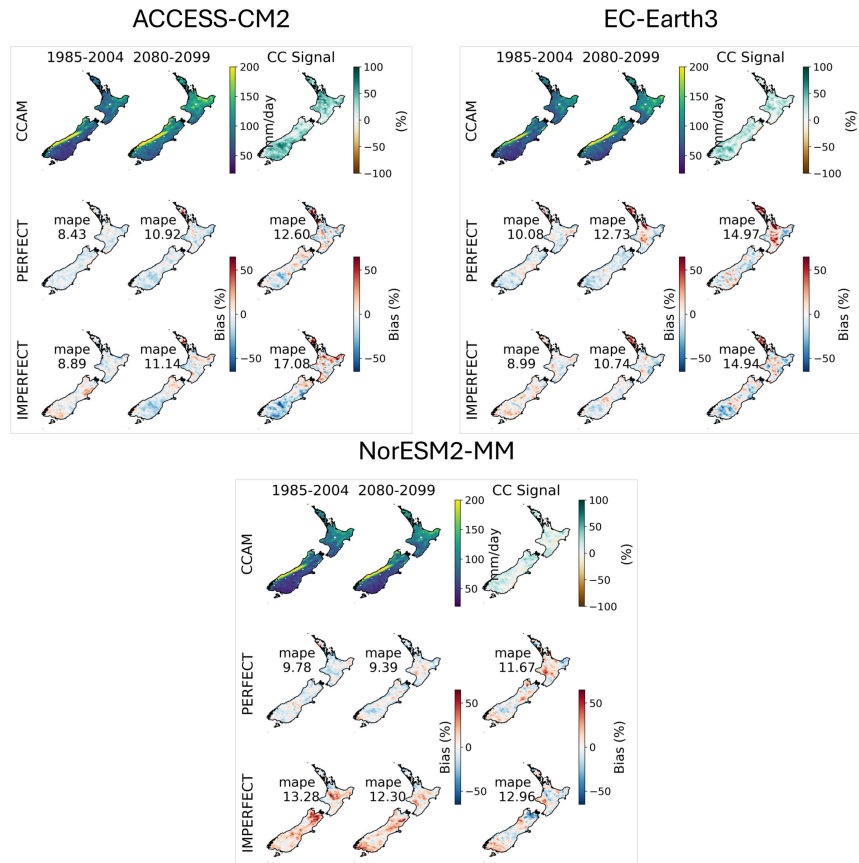

**Supplementary Figure S35:** Same as Figure S34, but for RX1Day (wettest-day-of-the-year).

# Further Sensitivity Testing of the Emulator (Figures S36-S38)

# Sensitivity Experiments for Emulator Training Setup

To better understand how the selection and quantity of GCM/RCM pairs used for training impact emulator performance, we conducted several sensitivity experiments. We used a simpler emulator (1.5 million parameters vs. 3.5 million in the main study) to enable faster training, with these sensitivity tests playing a key role in selecting the final model architecture and offering a more efficient alternative to full-scale testing.

These experiments aimed to:

- Confirm that our main findings were robust and were not due to a specific training setup or random chance.

- Determine whether training on multiple GCMs yields better performance than training on individual GCMs.

Specifically, we trained emulators on:

- Single GCMs (ACCESS-CM2, EC-Earth3, or NorESM2-MM individually) – 3 experiments for each variable.

- Two pairs of GCM (for example, ACCESS-CM2 and NorESM2-MM) – 3 experiments for each variable to cover all unique permutations of the GCM training pairs.

- Five or six GCMs – 1 experiment.

All emulators were trained for 200 epochs. The approximate training times using two A100 GPUs were: 15 hours for a single GCM, 30 hours for two GCM, and 75 hours for five GCMs. For multi-GCM training, each batch was stratified to include an equal number of training instances from each GCM.

For each experiment and variable, performance was assessed in-sample (on the training GCMs) and out-of-sample (on unseen GCMs), using both the perfect and imperfect frameworks. The training procedure for both precipitation and temperature followed the same approach described for the temperature model in the main manuscript, consistent with Doury et al. (2022).

To assess the differences between perfect and imperfect training performance, we summarize their differences across all experiments, spanning various GCMs, training setups, and metrics. Each point in the violin plots represents one emulator evaluated on three GCMs in both frameworks. Overall, imperfect performance is only slightly worse than perfect, with minimal differences for signal metrics, indicating that the emulator largely preserves the climate change signal (Figure S36). As shown in Figure S37, training on five GCMs did not improve future performance (2080–2099 climatology) compared to training in a single GCM, in the perfect or imperfect framework.

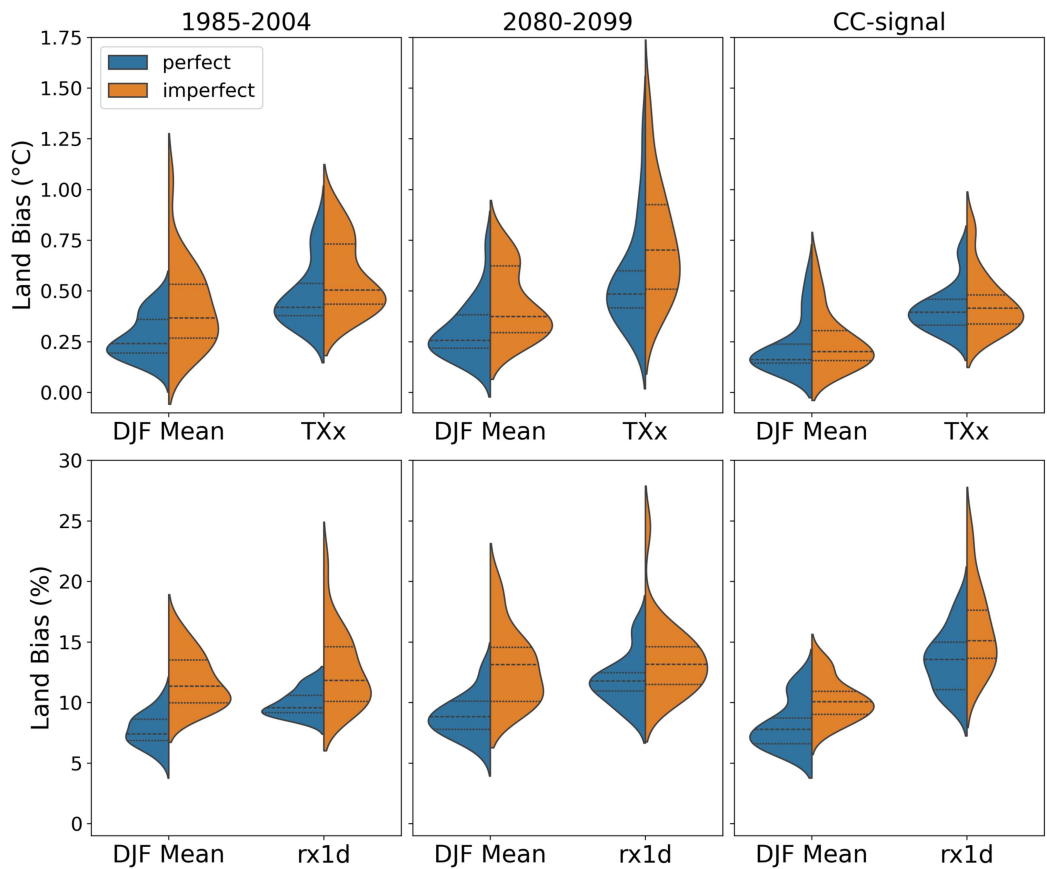

**Supplementary Figure S36**: Area-averaged RMSE for temperature (top row) and MAPE for precipitation (bottom row) for the historical period (first column), future period (second column), and climate change signal (third column). Results are shown for both the perfect (blue) and imperfect (orange) frameworks. The spread in the violin plots reflects performance across different GCMs and training setups (e.g., training on EC-Earth3 vs. ACCESS-CM2).

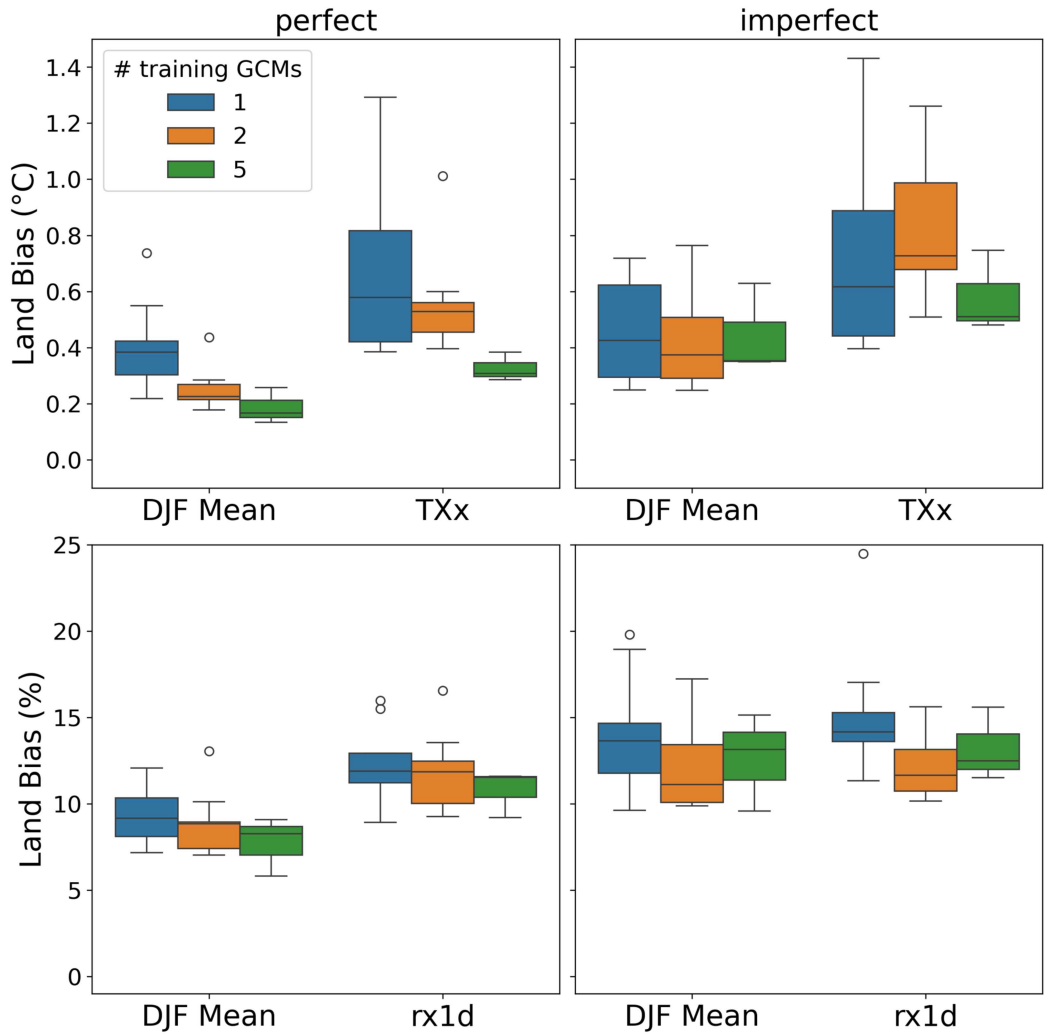

**Supplementary Figure S37**: Area-averaged RMSE for temperature (top row) and MAPE for precipitation (bottom row) over the future period (2080–2099), comparing emulator performance when trained on 1, 2, or 5 GCMs. The spread reflects variation across GCMs and experimental setups.

# Selection of GCMs (Figure 38)

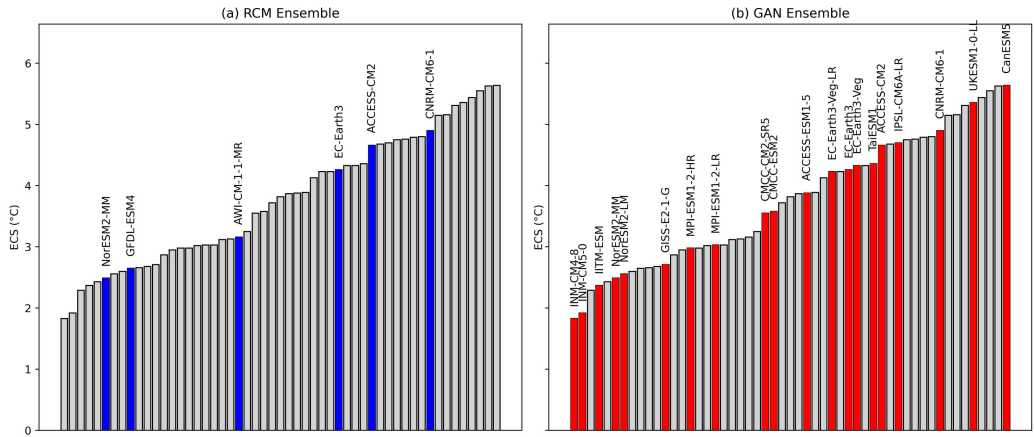

**Supplementary Figure S38**: Selected GCMs for the RCM ensemble (a) and the emulator (GAN) ensemble (b), shown with their equilibrium climate sensitivity, based on the 53 CMIP6 GCMs.

# Methods concerning uncertainty decomposition (Figures S39-S40)

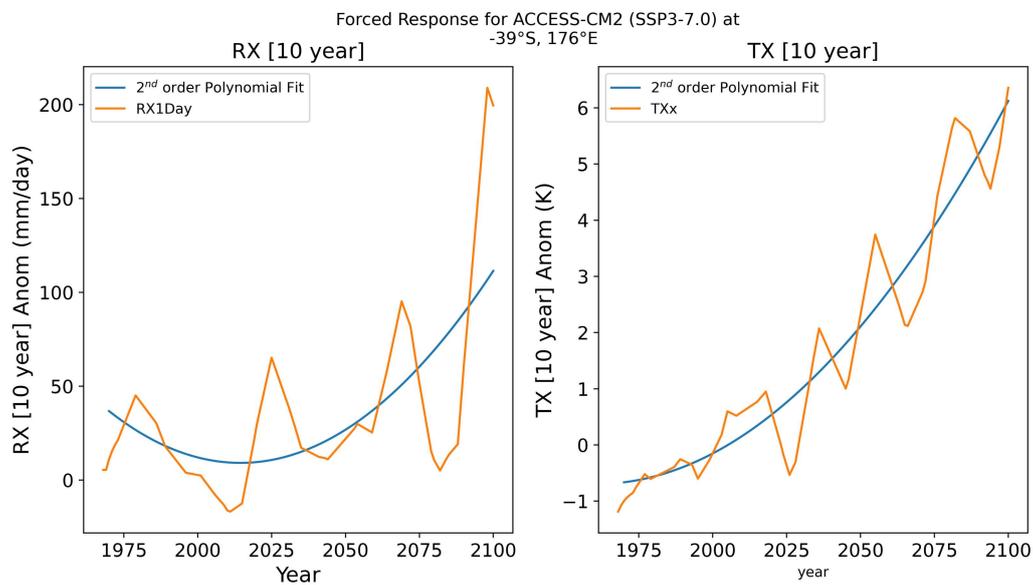

**Supplementary Figure S39**: Polynomial fits (2nd order) for RX [10 year] and TX [10 year].

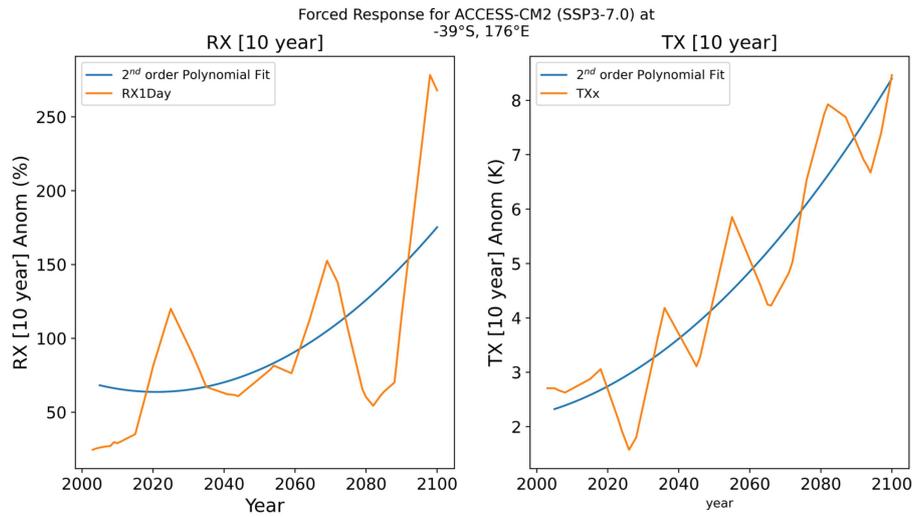

(a)

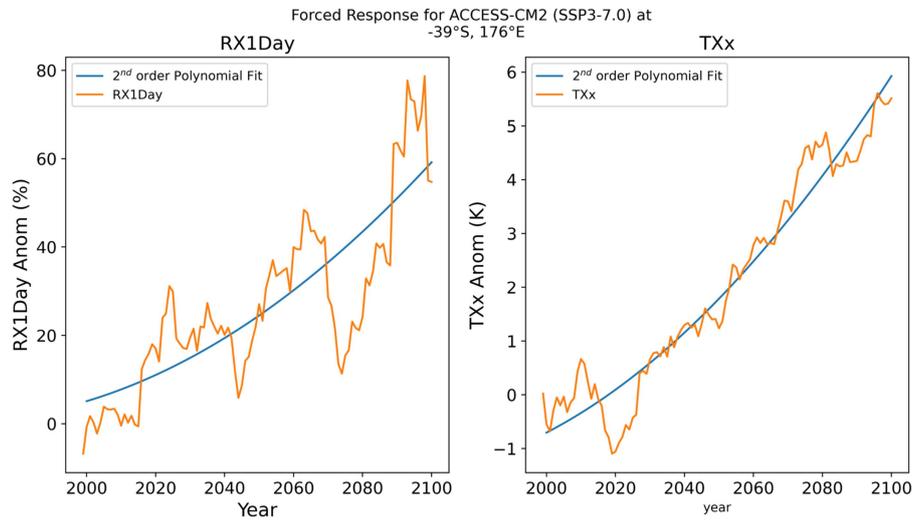

(b)

**Supplementary Figure S40**: Same as S39 but using percentage anomalies for precipitation (Figure S28). RX10yr and TX10yr appear smoother due to the use of rolling decadal maxima followed by a rolling average. RX1Day is also smoothed using a 10-year rolling average in both the top and bottom rows.

41



\end{document}